\documentclass[floatfix]{revtex4}
\usepackage{amsmath,amssymb,eucal,mathrsfs}
\usepackage[dvips]{epsfig,graphics}
\usepackage{graphicx}
\usepackage{amsmath}

\newcommand{\be}{\begin{equation}}
\newcommand{\ee}{\end{equation}}

\begin{document}

\title{Cold Dark Matter Hypotheses in the MSSM}  

\author{S.S. AbdusSalam}
\email{shehu@ictp.it}
\affiliation{Abdus Salam ICTP, Strada Costiera 11, I-34014 Trieste, Italia}

\author{F. Quevedo}
\email{f.quevedo@damtp.cam.ac.uk}
\affiliation{Abdus Salam ICTP, Strada Costiera 11, I-34014 Trieste,
  Italia \\ and DAMTP, Centre for Mathematical Sciences, Wilberforce
  Road, Cambridge CB3 0WA, UK} 


\begin{abstract}
We perform a Bayesian model selection analysis in the the R-parity
conserving MSSM to compare
two different assumptions: whether the lightest neutralinos make all or
only part of the cold dark matter. This corresponds to either imposing
full WMAP relic density limits or just its upper bound for
constraining the MSSM parameters. We consider several realisations of
the MSSM, namely, three GUT-scale SUSY breaking scenarios with a
handful of parameters corresponding to the CMSSM, anomaly mediation
and the large volume string scenarios as well as the weak-scale
25-parameter phenomenological MSSM (pMSSM). The results give a
data-based quantitative evidence for a multicomponent cold dark
matter. The pMSSM posterior samples indicate that the choice of
imposing full WMAP limits or just its upper bound affects mostly the
gaugino-higgsino content of the neutralino and, against naive
expectations, essentially not any other sector. 
\end{abstract}

\maketitle

Astrophysical observations~\cite{wmap7} indicate that more than $20\%$ of the
energy density of the universe is made of neutral and weakly
interacting non-baryonic `dark
matter'(DM) (for reviews, see Refs.
~\cite{Jungman:1995df,Bertone:2004pz,Garrett:2010hd}).  
Neutrinos are known to make up a minor component of the total
DM~\cite{arXiv:0803.0547}. The bulk of the DM has to be cold in order to be consistent with
structure formation in the universe. A cold DM (CDM) candidate 
should be massive and stable on cosmic time-scales. The neutralino lightest
supersymmetric particle (LSP) in the minimal supersymmetric standard
model (MSSM) with conserved R-parity is an excellent CDM
candidate.  Other CDM candidates from models beyond the
standard model  include axions/axinos,
gravitinos, Kaluza-Klein gravitons and stable moduli fields.

In most of the MSSM literature the neutralino has been considered to be 
the only thermal relic contribution to the DM energy density. However this assumption may be too strong and
unnecessary. In fact it is very difficult, i.e. requires much
fine-tuning, to obtain the observed value of the CDM relic
density with the above mentioned assumption. Typically 
the neutralino relic density  either radically
over-closes or under-closes the universe. For example bino-like
neutralinos yield relic densities 
between $1$ to $4$ orders of magnitude more than the WMAP fit value,
and it is known that in the minimal anomaly 
mediated symmetry breaking (mAMSB), which maps to a subset of the
MSSM parameters space, neutralino co-annihilations are so 
important that the predicted relic densities are generically far below
the cosmological and astrophysical fit value~\cite{Giudice:1998xp} to
the extent that an extra CDM component or some other modifications is
unavoidable~\cite{Baer:2010kd}. Mixed axion-axino~\cite{Baer:2010wm}
and/or stable moduli~\cite{arXiv:0705.3460} could make up additional
components to the neutralino CDM. In 
Refs.~\cite{Zurek:2008qg,Cholis:2009va,Feldman:2010wy} multicomponent 
DM models were constructed to explain its indirect detection
observations.  
 
The assumption that CDM are solely made of neutralinos and the
alternative case where additional non-neutralino components are
allowed are mutually exclusive hypotheses that could be compared
against each other in light of the currently available indirect
collider and astrophysical data. One could ask the question of which
of the two alternative hypotheses has more support from current data
relative to the other \footnote{This question is different
  from attempts as in Ref.~\cite{Buchmueller:2008qe} seeking to
  estimate the effects, on preferred CMSSM regions, of changing the
  error bars on the (solely neutralino) CDM relic density constraint
  from WMAP.}. Answers to questions of this sort, which can be updated
as new data become available, can be obtained via the Bayesian technique
for model selection.   

Bayesian model selection techniques have been successfully  applied in
cosmology and astrophysics, see for instance 
Ref.~\cite{astro-ph/0504022} and references therein, and recently in
particle physics phenomenology 
as well. For example the sign of the MSSM Higgs doublets mixing
parameter, $\mu$, is not yet fixed by any observation and is an
important quantity in determining possible SUSY contribution to the
muon anomalous magnetic moment, $\delta
a_\mu$~\cite{vonWeitershausen:2010zr}. In 
Ref.~\cite{0807.4512,AbdusSalam:2009qd} Bayesian technique  
were employed to select between the $sign(\mu) > 0$ and $sign(\mu) <
0$ branches of the constrained minimal supersymmetric model (CMSSM) 
and of the pMSSM. Another example which is 
conceptually different is that of Bayesian comparisons between
different models as illustrated in Ref.~\cite{AbdusSalam:2009tr} where 
various models for SUSY breaking mediation were compared with one
another in light of current indirect collider and WMAP data.  

\begin{figure}[t]
  \begin{tabular}{cc}
  (a) & (b) \\
   \begin{minipage}[t]{7.5cm}
    \includegraphics[width=1.\textwidth, height=.27\textheight]{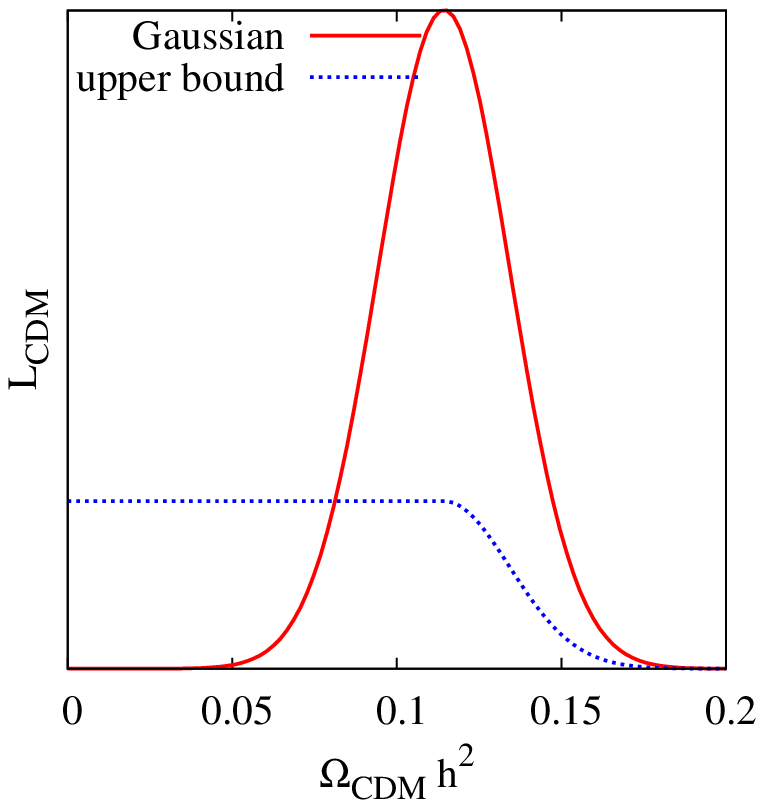}
   \end{minipage}
&
  \begin{minipage}[t]{9.5cm}
    \includegraphics[width=1.2\textwidth,
    height=.27\textheight]{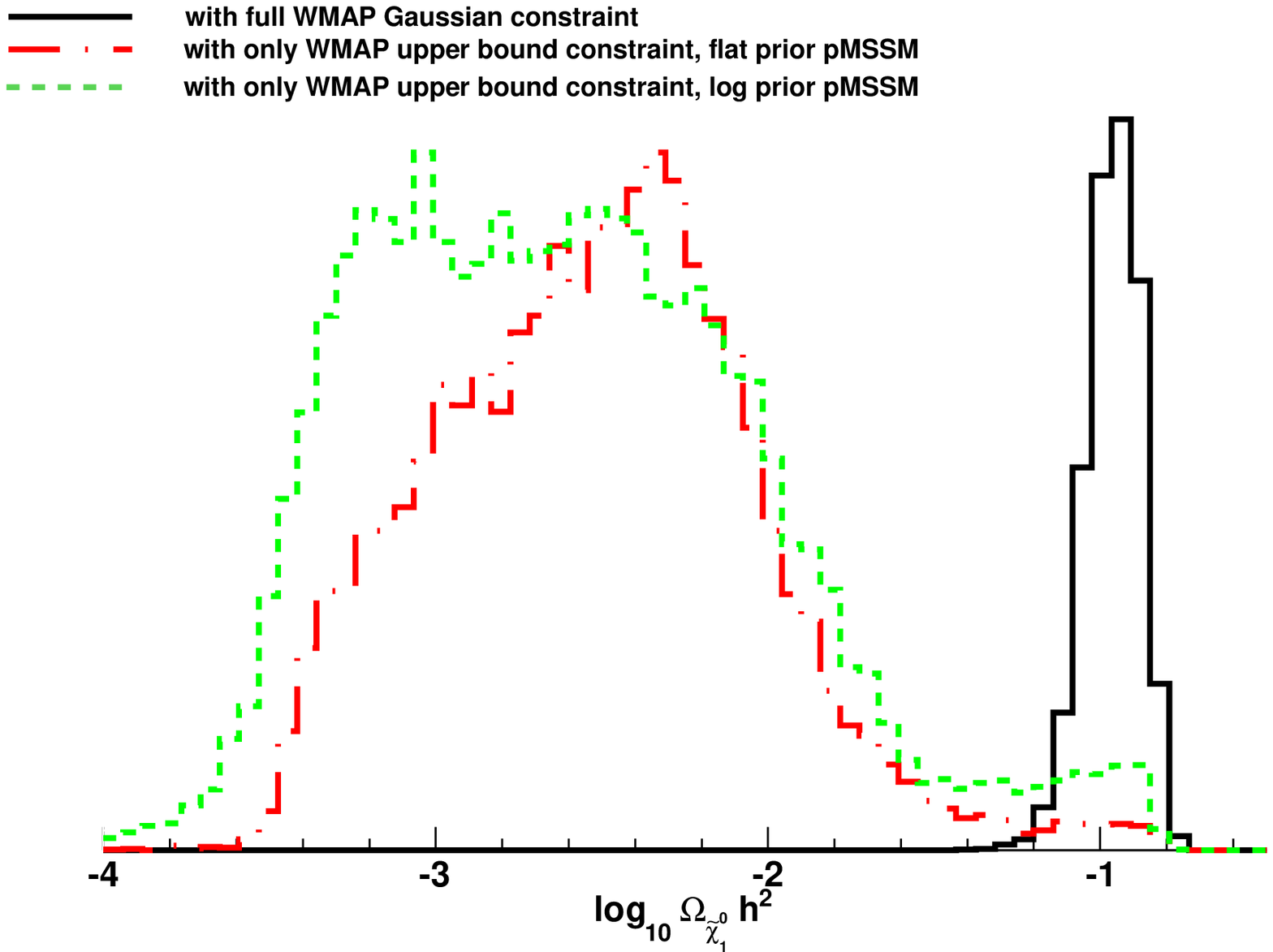}
   \end{minipage}
   \end{tabular}
  \caption{{\bf (a)}: The constraints put on pMSSM predicted values
  of $\Omega_{\mathrm{CDM}}h^2$ for the case where CDM are entirely due
  to the lightest neutralinos (solid line) or where a multicomponent
  CDM is allowed (dotted line). {\bf (b)}: The neutralino CDM
  relic density posterior distribution for ${\cal{H}}_0$ pMSSM hypothesis
  (solid line) and for ${\cal{H}}_1$ pMSSM hypothesis (dashed line.) For
  comparison the posterior from Ref.~\cite{AbdusSalam:2009qd} is also
  shown (dash-dotted line.) The posterior distributions show that
  generically the most preferred neutralino relic densities are
  centred around a point about two orders of magnitude less than the
  WMAP central value except when the single neutralino CDM scenario is
  imposed. This indicates that the single neutralino hypothesis is
  disfavoured. See Table II for a quantitave
  measure. } \label{fig.omegaboth}   
\end{figure}

Bayesian approach to MSSM and DM phenomenology for the LHC have been
used by various groups; see for instance \cite{Allanach:2006jc,
  Cabrera:2008tj, Raklev:2009uu, White:2010jp} and references
therein. In this article we compare the two CDM hypotheses in the
context of the pMSSM~\cite{Djouadi:1998di, AbdusSalam:2008uv,
  AbdusSalam:2009qd} and GUT-scale models for SUSY breaking
mediations, using Bayes' theorem and current data, to make 
statement about which of the hypotheses receive better support from
the data. In the first hypothesis, ${\cal{H}}_0$, the WMAP inferred
CDM data is imposed on the pMSSM and other GUT-scale models
parameters in a way compatible with 
the assumption that the whole CDM in the universe are solely made of
neutralinos. For the second hypothesis, ${\cal{H}}_1$, where a
multicomponent CDM is assumed, pMSSM and other GUT-scale models points
 that give relic 
densities lower than WMAP central point are not penalised so that
non-neutralino CDM components are allowed for. The way in which the 
WMAP CDM data is employed for the analysis is summarised in
Fig.~\ref{fig.omegaboth}(a). 

The remaining part of the article is
organised as follows. We first give a brief review of the 
Bayesian model selection technique followed by a description of how it
is applied for the pMSSM and other GUT-scale models for the two CDM hypotheses.
 Then we compare and explain some 
differences between the sampled points, particularly in the gaugino and
neutralino sectors, from this analysis and the previous corresponding
pMSSM analysis considered in Ref.~\cite{AbdusSalam:2009qd}. Finally we
explain the implications of the hypotheses for direct DM detection 
experiments before ending the article with summary and conclusions. 

Here we give definitions and set terminology 
that would allow for a self-contained discussion. Bayesian inference
studies are centered around two quantities: model evidence and
parameters posterior probability distributions. These are defined as
follows. Consider a hypothesis (or model with all its assumptions) $\cal{H}$
with parameters 
$\underline m$. The {\it a priori} assumed form of values which the model
parameters can take is encoded in the prior probability distribution
of the model parameters given that the hypothesis is true, $p(\underline m | {\cal{H}})$. 
The posterior distribution of the parameters in light of the data
considered and
in the context of the hypothesis is represented by $p(\underline m|
\underline d,{\cal{H}})$. The data, $\underline d$, is a set of
experimentally determined values for certain quantities (observables) 
which the model can predict. The support or evidence which the
model obtains from the data is defined as the
probability density of observing the data set given that the
hypothesis is true and represented by $\mathcal{Z} \equiv p(\underline
d|{\cal{H}})$.   

The evidence is calculated as  
\begin{equation} \label{eq.evid}
\mathcal{Z} =
\int{p(\underline d|\underline m,{\cal{H}}) p(\underline m|{\cal{H}})}\ d \underline m
\end{equation}
where the integral is $N$-dimensional, with $N$ the dimension of the
set of parameters 
$\underline m$. $p(\underline d|\underline m, {\cal{H}})$ is called 
the likelihood, it is the probability 
of obtaining the data set $\underline d$ from certain model
parameters $\underline m$ and is a 
function of $\chi^2$ for the analyses presented in this
article. Eq.(\ref{eq.evid}) is obtained directly from 
Bayes' theorem 
\begin{equation} \label{eq.bayes}
p(\underline m|\underline d, {\cal{H}}) = \frac{p(\underline d|\underline
  m,{\cal{H}})p(\underline m|{\cal{H}})}{p(\underline d|{\cal{H}})}.
\end{equation}
In the reign of searches for new physics~\cite{Nath:2010zj} from
upcoming LHC and other experimental data, Bayesian inference analysis
would play an important role in selecting between various models based
on the data. For such $LHC^{-1}$ problem and the Bayesian selection
example presented in this article, evaluating the evidence is more
important compared to computing parameters posterior probability distributions. In
general one would not bother about finding a model posterior
probability distribution if, to begin with, it is known that its evidence is
poor. However, evaluating the evidence is computationally very
expensive especially for multi-dimensional parameter spaces. In fact
until the work in Ref.~\cite{Skilling}, the magnitude of the evidence
were mostly disregarded and only the model posterior probability
distributions were computed. For evaluating the evidence we use
the MultiNest algorithm~\cite{Feroz:2008xx,Feroz:2007kg} which implements nested
sampling technique~\cite{Skilling} and is very efficient in dealing with
complex and multi-modal parameter spaces.  

In order to select between two hypotheses ${\cal{H}}_{0}$ and ${\cal{H}}_{1}$
one needs to compare their respective posterior probabilities given
the data set $\underline d$ to be used for the comparison. The
posterior probabilities are compared via Bayes' theorem as   
\begin{equation} \label{eq:3.1}
\frac{p({\cal{H}}_{1}|\underline d)}{p({\cal{H}}_{0}|\underline d)}
=\frac{p(\underline d|{\cal{H}}_{1})p({\cal{H}}_{1})}{p(\underline d|
{\cal{H}}_{0})p({\cal{H}}_{0})}
=\frac{\mathcal{Z}_1}{\mathcal{Z}_0}\frac{p({\cal{H}}_{1})}{p({\cal{H}}_{0})},
\end{equation}
where $p({\cal{H}}_{1})/p({\cal{H}}_{0})$ is the ratio of prior probabilities for the
two models usually set to unity with the assumption that the two
hypothesis are {\it a priori} equally likely. $\mathcal{Z}_1/\mathcal{Z}_0 >
1$ would mean that the data favours 
model ${\cal{H}}_1$ relative to ${\cal{H}}_0$ and vice versa if the ratio is less than
one. Given various evidences to be compared beside one another,
Jeffreys' scale~\cite{Jeffreys}, which gives a calibrated spectrum of
significance for the relative strength between the evidences, 
\begin{equation} \label{eq:Jeffreys}
\Delta \log_e \mathcal{Z} = \log_e \left[ \frac{p({\cal{H}}_{1}|\underline
    d)}{p({\cal{H}}_{0}|\underline d)}\right] = \log_e \left[
    \frac{\mathcal{Z}_1}{\mathcal{Z}_0}\right],
\end{equation}
is used. The Jeffreys' scale convention we employ is shown in 
Tab.~\ref{tab:Jeffreys}. Values in the column labeled
``Probability'' are calculated as follows. Given that ${\cal{H}}_0$ and ${\cal{H}}_1$
are mutually exclusive and {\it a priori} equally likely then  
\begin{equation}
p({\cal{H}}_0|\underline d) + p({\cal{H}}_1|\underline d) = 1 \textrm{ and } p({\cal{H}}_0)= 
p({\cal{H}}_1) 
\end{equation}
so that the posterior probabilities are given by 
\begin{equation}
p({\cal{H}}_0|\underline d) = \frac{\mathcal{Z}_0}{\mathcal{Z}_0 +
  \mathcal{Z}_1} \textrm{ and } p({\cal{H}}_1|\underline d) =
\frac{\mathcal{Z}_1}{\mathcal{Z}_0 + \mathcal{Z}_1}. 
\end{equation}
\begin{table}
\begin{tabular}{|ccc|}
\hline
$|\Delta \log_e \mathcal{Z}|$ & Probability & Remark \\ 
\hline
$<1.0$ & $<0.750$ & Inconclusive \\
$1.0$ &  $0.750$ & Weak Evidence \\
$2.5$ &  $0.923$ & Moderate Evidence \\
$5.0$ &  $0.993$ & Strong Evidence \\ \hline
\end{tabular}
\caption{Jeffreys' scale for the interpretation of relative
  evidences. $\Delta \log_e \mathcal{Z} > 5$, e.g. 10, would indicates
  an irrefutable relative evidence.} 
\label{tab:Jeffreys}
\end{table}
Next, we are going to apply the selection technique presented above to
assess the mutually exclusive CDM hypotheses using current indirect
collider and WMAP data. 

The pMSSM~\cite{Djouadi:1998di, AbdusSalam:2008uv, Berger:2008cq, AbdusSalam:2009qd} has unconstrained
MSSM parameters defined at the weak-scale. In order to suppress sources
of unobserved 
CP-violation and FCNC the parameters are set to be real with
diagonal sfermion mass parameters, $m_{\tilde{f}}$, and trilinear
scalar couplings. In addition, first and second generation mass
parameters are 
set to be degenerate. The soft SUSY breaking parameters that remain, 
out of the initially more than 100 parameters, make the pMSSM parameters set
\be 
\underline m = \{\tan \beta, \; m^2_{{{H}}_1}, \; m^2_{H_2}; \;                                      
       M_{1,2,3};\; m^{3rd gen}_{\tilde{f}_{1,2,3,4,5}},\; m^{1/2nd
         gen}_{\tilde{f}_{1,2,3,4,5}}; A_{t,b,\tau,\mu=e} \}
\ee
where the notations are as in Ref.~\cite{AbdusSalam:2009qd} and
the integers $1,2,3,4,5$ label third generation and degenerate first
and second generation of the sfermion masses there. 

Cosmological and astrophysical fits to the standard cosmological
constant plus CDM model imply a relic density \footnote{Here we use
  the five years WMAP values in order to properly compare with
  Ref.~\cite{AbdusSalam:2009qd}. Updating this to the seven years WMAP value would not change
  the main results and conclusions of this work.} 
\be \label{drelic} \Omega_{CDM} h^2 = 0.1143 \pm 0.0034 \ee 
where $h$ is the reduced Hubble constant~\cite{arXiv:0803.0547}. 
Assuming that the thermal relic of the LSP 
account for the magnitude in Eq.(\ref{drelic}), the 
${\cal{H}}_0$ hypothesised pMSSM predictions for the relic density are
constrained to lie within the observed central value but with an
inflated error to cater for uncertainties in theoretical
predictions~\cite{Baro:2007em}   
\be \label{dmcons} \Omega_{CDM} h^2 = 0.1143 \pm 0.02.\ee 
On the other hand, for ${\cal{H}}_1$ only the upper half of the above constraint,
Eq.(\ref{dmcons}), is imposed such that predicted relic densities
lower than the central value are not penalised as shown in
Fig.~\ref{fig.omegaboth}(a). 

Apart from the CDM relic we also impose precision electroweak and
B-physics indirect collider data for checking the neutralino CDM
hypotheses. Various predictions, $\underline O$, for the observables
were obtained via \texttt{SOFTSUSY2.0.17}~\cite{Allanach:2001kg} for
producing the MSSM spectrum;
\texttt{micrOMEGAs2.2}~\cite{Belanger:2008sj} for computing the
neutralino CDM relic density, the branching ratio $BR(B_s \rightarrow
\mu^+ \mu^-)$ and the anomalous magnetic moment of the muon
$(g-2)_\mu$; \texttt{SuperIso2.0}~\cite{Mahmoudi:2007vz} for
predicting the Isospin asymmetry in the decays $B \rightarrow K^*
\gamma$ and $BR(B \rightarrow s \gamma)$; 
and
\texttt{susyPOPE}~\cite{Heinemeyer:2006px,Heinemeyer:2007bw} for
computing precision observables that include the $W$-boson mass $m_W$,
the effective leptonic mixing angle variable $\sin^2
\theta^{lep}_{eff}$, and the total $Z$-boson decay width, $\Gamma_Z$,
at two loops in the dominant MSSM parameters. The experimentally
determined central values, $\underline \mu$, and 
their corresponding uncertainties, $\underline \sigma$, of the
observables  
\begin{eqnarray} 
\underline O &= &\{ m_W,\; \sin^2\, \theta^{lep}_{eff},\; \Gamma_Z,\; \delta
a_{\mu},\; R_l^0,\; A_{fb}^{0,l},\; A^l = A^e,\; R_{b,c}^0,\;
A_{fb}^{b,c},\; A^{b,c}, \\ \nonumber 
& &BR(B \rightarrow X_s \, \gamma),\; BR(B_s \rightarrow \mu^+ \, \mu^-),\; 
\Delta_{0-},\; R_{BR(B_u \rightarrow \tau \nu)},\; R_{\Delta M_{B_s}},\\ \nonumber 
& &\Omega_{CDM}h^2 \} 
\end{eqnarray}
make the data set (same as those in
Ref.~\cite{AbdusSalam:2009qd}) for the analysis 
\be \label{dat} \underline d = \{ \mu_i, \sigma_i \} \ee where $i = 1,
\ldots, 19$ 
label the individual observables. The central values and errors of the
observables are summarised in Tab.~\ref{tab:obs}. 
\begin{table}
\begin{center}{\begin{tabular}{|clll||clll|}
\hline
Observable & Constraint & Th. Source & Ex. Source & Observable &
Constraint & Th. Source & Ex. Source  \\ 
\hline
$m_W$ [GeV]& $80.399 \pm  0.027$& \cite{Heinemeyer:2006px},
\cite{Heinemeyer:2007bw}& \cite{verzo}  &$A^l = A^e$& $0.1513 \pm
0.0021$  & \cite{spope} & \cite{:2005ema} \\
$\Gamma_Z$ [GeV]& $2.4952 \pm 0.0025$& \cite{spope} &  \cite{:2005ema} 
&$A^b$ & $0.923 \pm 0.020$ & \cite{spope} & \cite{:2005ema}  \\
$\sin^2\, \theta_{eff}^{lep}$  & $0.2324 \pm 0.0012$& \cite{spope} &
\cite{:2005ema} &$A^c$ & $0.670 \pm 0.027$ & \cite{spope} &
\cite{:2005ema} \\  
$\delta a_\mu $ & $(30.2 \pm 9.0) \times 10^{10}$& 
\cite{Moroi:1995yh,Degrassi:1998es,Heinemeyer:2003dq,Heinemeyer:2004yq}
& \cite{Bennett:2006fi,Davier:2007ua, Hertzog:2007hz} &$Br(B
\rightarrow X_s \gamma)$ & $(3.55 \pm 0.42) \times 10^{4}$ & 
\cite{Misiak:2006zs, Misiak:2006bw, Misiak:2006ab, Becher:2006pu} &
\cite{Barberio:2007cr} \\  
$R_l^0$ & $20.767 \pm 0.025$ & \cite{spope} & \cite{:2005ema} &$Br(B_s
\rightarrow \mu^+ \mu^-)$ & see Fig.~\protect~\ref{fig.bsmumu}& 
\cite{Belanger:2008sj,Belanger:2006is,Belanger:2004yn,Belanger:2001fz} 
& \cite{:2007kv}  \\ 
$R_b^0$ & $0.21629 \pm 0.00066$& \cite{spope} & \cite{:2005ema}
&$R_{\Delta M_{B_s}}$ & $0.85 \pm 0.11$ & \cite{Bona:2006ah} &
\cite{Abulencia:2006ze}      \\ 
$R_c^0$ & $0.1721 \pm 0.0030$& \cite{spope} & \cite{:2005ema}
&$R_{Br(B_u \rightarrow \tau \nu)}$ & $1.26 \pm 0.41$&
\cite{Isidori:2006pk,Isidori:2007jw,Akeroyd:2003zr} &
\cite{Aubert:2004kz,paoti,hep-lat/0507015}      \\ 
$A_{\textrm{FB}}^b$ & $0.0992 \pm 0.0016$& \cite{spope} &
\cite{:2005ema} &$\Delta_{0-}$ & $0.0375 \pm 0.0289$&
\cite{Mahmoudi:2007vz} & \cite{J.Phys.G33.1} \\ 
$A_{\textrm{FB}}^c$ & $0.0707 \pm 0.035$ & \cite{spope} &
\cite{:2005ema} &$\Omega_{CDM} h^2$ & $0.11 \pm 0.02 $&
\cite{Belanger:2008sj,Belanger:2006is,Belanger:2004yn,Belanger:2001fz}
& \cite{0803.0547} \\
\hline
\end{tabular}}\end{center}
\caption{Summary for the central values and errors for the
  observables. Theoretical uncertainties have been added in quadrature
  to the experimental uncertainties quoted.} 
\label{tab:obs}
\end{table}
 \begin{figure}
     \centering\includegraphics[width=.5\textwidth]{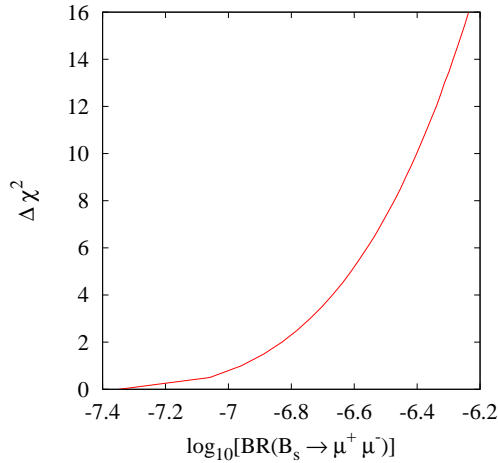}
   \caption{The chi-squared contour used for the B-physics observable
     $BR(B_s \rightarrow \mu^+ \mu^-)$.}   
  \label{fig.bsmumu} 
 \end{figure}
We assume that the observables are independent to form combined
likelihoods   
\be p(\underline d|\underline m, {\cal{H}}_0) = \prod_{i=1}^{19} \,
\frac{ \exp\left[- (O_i - \mu_i)^2/2 \sigma_i^2\right]}{\sqrt{2\pi
    \sigma_i^2}} \textrm{ and } 
 p(\underline d|\underline m, {\cal{H}}_1) = L(x) \prod_{i=1}^{18} \,
\frac{ \exp\left[- (O_i - \mu_i)^2/2 \sigma_i^2\right]}{\sqrt{2\pi
    \sigma_i^2}}
\ee   
where for ${\cal{H}}_1$ the index $i$ run over the different experimental observables 
(data) other than the CDM relic density; $x$ represents the
predicted value of neutralino CDM relic density;
\be \label{olik}
L(x) = 
\begin{cases}
1/(y + \sqrt{\pi s^2/2}), & \textrm{if $x < y$} \\  
\exp\left[-(x-y)^2/2s^2\right]/(y + \sqrt{\pi s^2/2}),
& \textrm{if $x \geq y$} \\
\end{cases};
\ee 
$y = 0.1143$ is the WMAP central value quoted in Eq.(\ref{dmcons}) and
$s=0.02$ the inflated error. $L(x)$, shown in
Fig.~\ref{fig.omegaboth}(a), is the 
likelihood function for ${\cal{H}}_0$ and ${\cal{H}}_1$ respectively corresponding to
the regions where $x < y$ and $x \geq y$. Next we present and discuss 
the results of the samplings in the paragraphs that follow. 

The Bayesian evidence values from the samplings are summarised in
Tab.~\ref{tab:factors}. 
\begin{table}
\begin{tabular}{|cccc|}
\hline 
Models & $\Delta \log_e \mathcal{Z}_{flat}$ & $\Delta \log_e
\mathcal{Z}_{log}$ & $P(CDM=\tilde{\chi}_1^0)$ \\  
\hline \hline
pMSSM~\cite{AbdusSalam:2008uv} & 3.2 & - & 0.04 \\
pMSSM~\cite{AbdusSalam:2009qd} & - & 2.7 & 0.06 \\
\hline \hline
CMSSM & 1.4 & 2.5 & 0.18 \\
mAMSB & 14.1 & 14.8 & $6.8 \times 10^{-7}$ \\ 
LVS  & 3.6 & 3.7 & 0.02 \\
\hline
\end{tabular}
\caption{$\Delta \log_e \mathcal{Z}_i$ values for the 
  mutually exclusive CDM hypotheses in the pMSSM and some high scale models
  of SUSY breaking mediation. $P(CDM=\tilde{\chi}_1^0)$ represents the
  probabilities that the CDM are solely made of neutralinos.  For the details of the comparisons between the
  CMSSM, mAMSB and the LARGE volume scenario
  (LVS)~\cite{Conlon:2006wz} models see
  Ref.~\cite{AbdusSalam:2009tr}.}   
\label{tab:factors}
\end{table}
Let us first comment on the case of the pMSSM. For the setting  of Ref.~\cite{AbdusSalam:2008uv},
where only the linear prior was used, we obtain the following evidences:
$\log_e \mathcal{Z}_0 = 29.902 \pm 0.040$, $\log_e \mathcal{Z}_1 =
33.120 \pm 0.028$ and difference $\Delta \log_e \mathcal{Z} = 3.218$. Wider 
prior parameters  ranges and more B-physics observables were added 
for the new samplings for this article and for those in
Ref.~\cite{AbdusSalam:2009qd}  
relative to those in Ref.~\cite{AbdusSalam:2008uv}; as such the
absolute evidence values differ significantly. However, here only the
difference in the evidence values are important for the purpose of
this paper. A significant difference in $\Delta \log_e \mathcal{Z}$
is not expected if for the flat prior the constraints applied are as
in Ref.~\cite{AbdusSalam:2009qd} in place of those in
Ref.~\cite{AbdusSalam:2008uv} from which the flat prior 
result quoted in Tab.~\ref{tab:factors} is obtained  \footnote{The main difference between
    the pMSSM analysis in Ref.\cite{AbdusSalam:2008uv} and
    Ref.\cite{AbdusSalam:2009qd} is that the latter have additional
    set of electroweak observables from the code
    \texttt{susyPOPE}~\cite{Heinemeyer:2006px,Heinemeyer:2007bw} and
    have extended parameter ranges up to 4 TeV compared to 2 TeV for
    the former.}.  
The log prior results in the pMSSM~\cite{AbdusSalam:2009qd} are
$\log_e \mathcal{Z}_0 = 65.043 \pm 0.042$, $\log_e \mathcal{Z}_1 =
67.761 \pm 0.030$ with $\Delta \log_e \mathcal{Z} = 2.718$. 
Translating the evidence ratios using Jeffreys' scale,
Tab.~\ref{tab:Jeffreys}, implies that current indirect collider and 
cosmological data as employed here shows a significant evidence in
support of multicomponent CDM hypothesis. The probability of an
entirely neutralino-made CDM is around $0.04$ or $0.06$ respectively
for the flat and log prior pMSSM. The pMSSM results imply an
approximate prior independence since changing the log prior to a flat
prior for the \cite{AbdusSalam:2009qd} fit should not lead to a large
difference in $\Delta \log_e \mathcal{Z}$.

 

In the second part of Tab.~\ref{tab:factors}, the difference
in the evidence values for the two hypotheses 
${\cal{H}}_0$ and ${\cal{H}}_1$,  is presented in the
context of CMSSM, mAMSB and the LARGE volume scenarios
(LVS)~\cite{Conlon:2006wz,AbdusSalam:2007pm}.  Using the
Jeffreys' scale Tab.~\ref{tab:Jeffreys} we can extract the following information: the CMSSM results are prior dependent
ranging from being weak to moderate evidence for the
multicomponent CDM hypothesis. The LVS is approximately prior
independent with a clear-cut indication of the presence of significant
evidence against the solely neutralino CDM hypothesis,
${\cal{H}}_0$. The mAMSB case is overwhelmingly prior independent and 
give an irrefutable evidence for multicomponent CDM hypothesis. 
The latter result can be considered as a check that proves that the
Bayesian model selection technique is working since it is 
known~\cite{Giudice:1998xp} before hand that the mAMSB neutralino CDM
cannot account for the WMAP fit, Eq.(\ref{dmcons}), and hence a
multicomponent CDM is necessary in that scenario.  

Note that from the evidence computations, the posterior samples of the
models considered are obtained for free. For instance the posterior
distributions of the predicted neutralino CDM for both hypotheses
${\cal{H}}_0$ and ${\cal{H}}_1$ are shown in 
Fig.~\ref{fig.omegaboth}(b). The distribution for ${\cal{H}}_1$ is centred at
a value far less than the observed central value showing that addition
non-neutralino CDM component(s) is(are) necessary in order to account
for the WMAP data. Other analysis we performed from the posterior
samples of the models are presented in the remaining parts of this
article. We present and compare the difference between samples here
for ${\cal{H}}_1$ and from Ref.~\cite{AbdusSalam:2009qd} for ${\cal{H}}_0$. We address
mostly the neutralino and chargino sectors which get most affected by
the choice of ${\cal{H}}_0$ or ${\cal{H}}_1$ hypotheses. 
 
The nature of the neutralino determines the processes by which it
annihilates and/or co-annihilates into standard model particles and therefore
the main determining factor for its relic density.  
The neutralino mass is given by $\frac{1}{2} 
   {\psi^0}^T M_N \psi^0 + H.c.$ where ${\psi^0}^T = ( - i \tilde{b},
   - i \tilde{w}^3,\tilde{{{H}}_1^0}, \tilde{H_2^0} )$ and,  
\be \label{neutmass}
M_\textrm{N} = \left( \begin{array}{cccc} 
M_1 & 0 & - m_Z c_\beta s_W & m_Z s_\beta c_W \\
0 & M_2 &  m_Z c_\beta c_W & - m_Z s_\beta c_W \\
- m_Z c_\beta s_W & m_Z c_\beta c_W &  0 & - \mu \\
 m_Z s_\beta s_W & - m_Z s_\beta c_W &  -\mu & 0 
\end{array} \right),
\ee
$c_x = \cos x$ and $s_x = \sin x$.
The neutralino mass eigenstates are $\tilde{\chi}_i^0 = N_{ij}
\psi_j^0$ where N is a unitary transformation that diagonalises
$M_N$. 
The neutralino mass eigenstate is a mixture of bino $\tilde{b}$, wino
$\tilde{w}^3$ and higgsinos $\tilde{H}_{1,2}$
\be \label{neut1}
\tilde{\chi}_1^0 = N_{11}\tilde{b} + N_{12}\tilde{w}^3 +
   N_{13}\tilde{{{H}}_1^0} + N_{14}\tilde{H_2^0}, \quad \sum_{i=1,2,3,4}
   (N_{1i})^2 = 1.
\ee
The coefficients $N_{1i}$ with $i=1,2,3,4$, and the neutralino masses
are complicated functions of the soft terms $M_1, M_2, \mu, sign(\mu),
\tan \beta$ etc (see e.g. \cite{ElKheishen:1992yv} and references
therein) however the following statements approximate some relations
between the $N_{1i}$s and the parameters. 
When $M_1 \ll min(M_2,|\mu|)$, $N_{11}$ dominates and 
the LSP is dominantly bino. Bino LSPs give relic densities too 
high, beyond the WMAP limits, in most regions of parameter space and
therefore any pMSSM point in both ${\cal{H}}_0$ and ${\cal{H}}_1$ priors with such
property would be excluded by the relic density constraint
Eq.(\ref{dmcons}). Next when $M_2 < min(M_1, |\mu|)$, $N_{12}$
dominates so 
the the neutralino is dominantly wino and is quasi mass degenerate 
with the lightest chargino. This leads to strong chargino
co-annihilations to the extent that the relic density is typically
much smaller than the WMAP constraint. The pMSSM parameter points that
have that type of neutralino would be excluded in the ${\cal{H}}_0$ prior
scenario but not in ${\cal{H}}_1$. For $|\mu| < min(M_1, M_2)$, 
$N_{13}$ and $N_{14}$ are of order one and the LSP is dominantly
higgsinos-like. In this case the neutralino efficiently annihilates
into top and weak gauge boson pairs. In addition 
the $\tilde{\chi}_1^0$, $\tilde{\chi}_2^0$, and $\tilde{\chi}_1^\pm$
are all approximately mass degenerate and higgsinos-like and hence
there are more open channels for co-annihilations. Parameter points
with this type of neutralino are also excluded in ${\cal{H}}_0$ prior.  

\begin{figure}[t]
    (a) \\
    \includegraphics[width=.4\textwidth,
    height=0.25\textwidth]{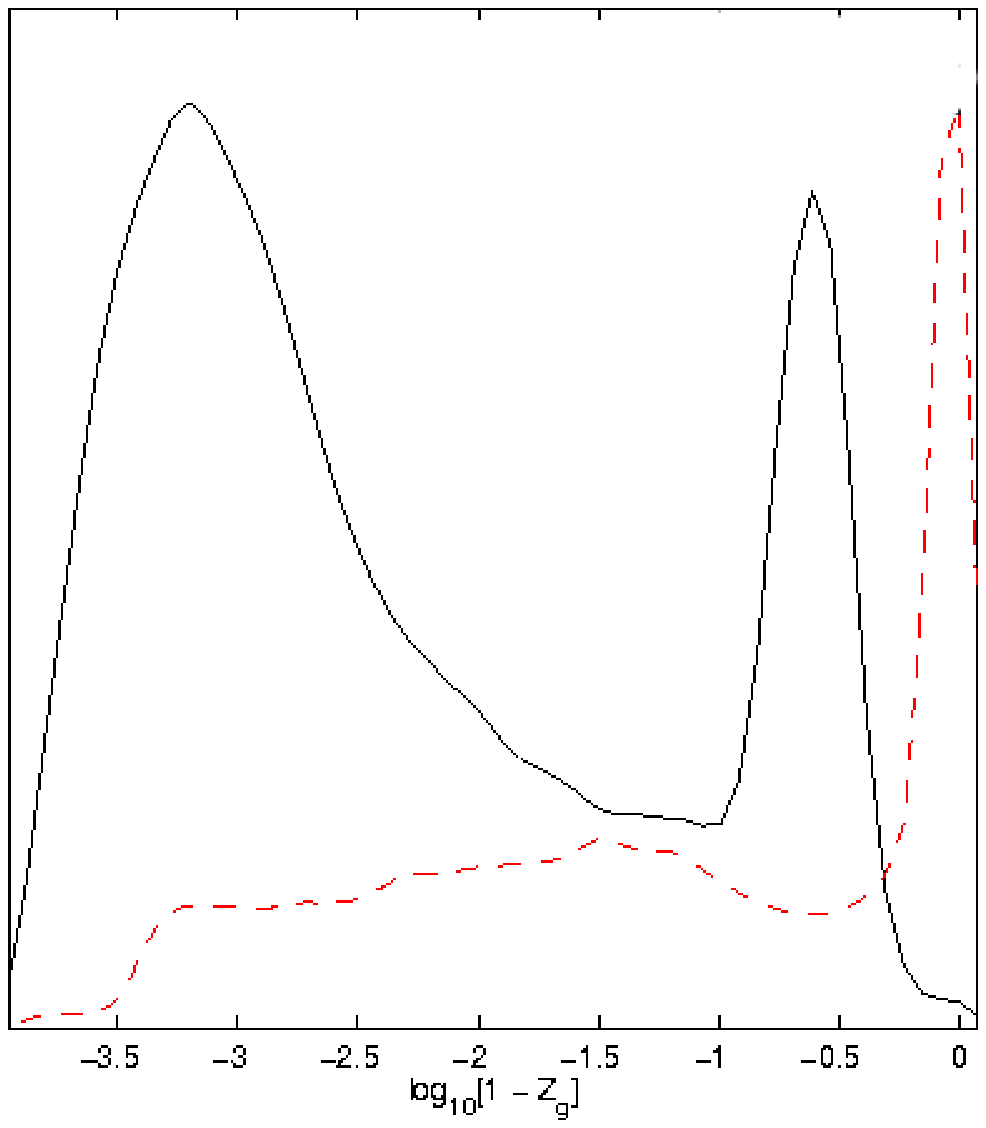} \\
  \begin{tabular}{cc}
    (b) & (c) \\
   \begin{minipage}[t]{8.5cm}
    \includegraphics[width=.9\textwidth, height=0.5\textwidth]{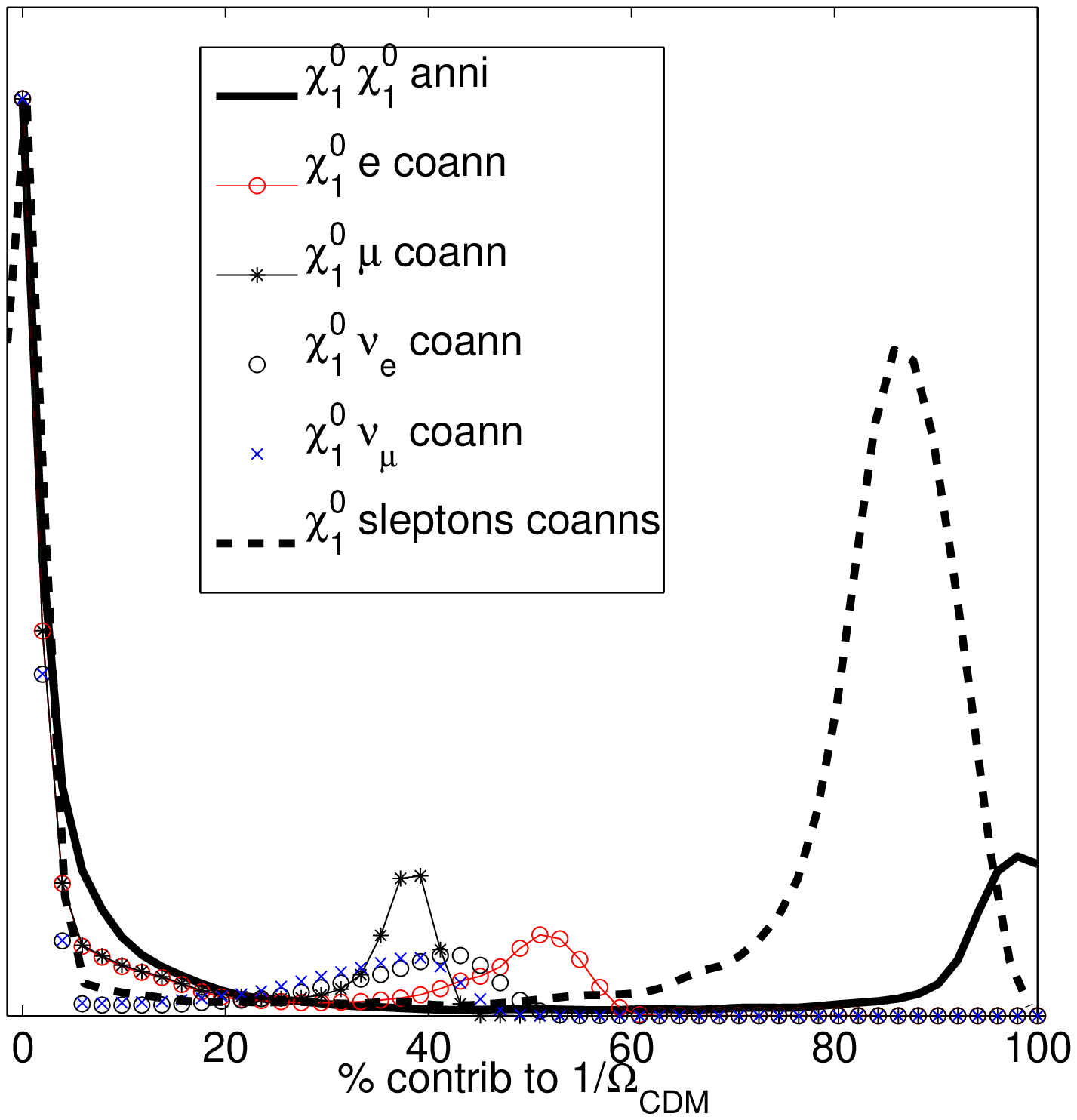}
   \end{minipage}
&
  \begin{minipage}[t]{8.5cm}
    \includegraphics[width=.9\textwidth, height=0.5\textwidth]{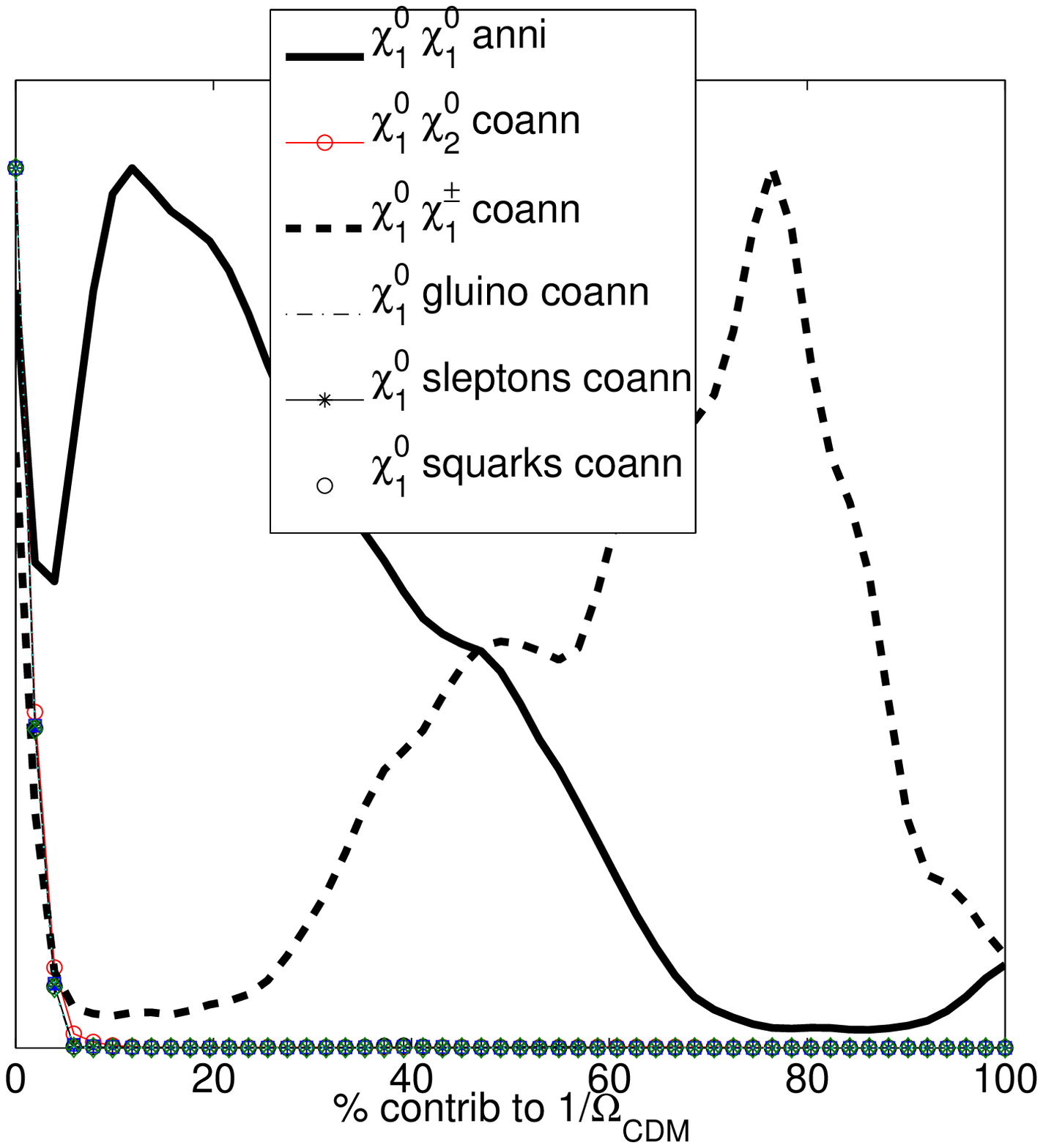}
   \end{minipage}
   \end{tabular}
  \caption{{\bf (a)} Neutralino's gaugino-higgsino admixtures for
    ${\cal{H}}_0$ (solid line) and for
    ${\cal{H}}_1$ (dashed line) log prior pMSSM CDM hypotheses. The
    neutralino is 
    almost always purely higgsino (peak around zero) for the
    ${\cal{H}}_1$ hypothesis due to a single efficient dominant
    co-annihilation that leads to posteriors with much lower (than
    WMAP central value) relic densities. For the ${\cal{H}}_0$
    hypothesis, however, there are mainly two dominant 
    (co-)annihilation channels that keep the relic densities within
    the imposed WMAP limit. With $Z_g = |N_{11}|^2 +
    |N_{22}|^2$, peaks to the right-hand side of the abscissa indicate
    higgsino domination. {\bf (b)} ${\cal{H}}_0$ prior pMSSM posterior sample
    distribution for the early universe neutralino annihilation (solid
    line) and co-annihilations whose cross-sections dominate in 
    determining the present neutralino relic
    density. Neutralino-sleptons co-annihilation is dominant in this  
    scenario since the neutralinos are mostly gaugino-like as shown in
    plot (a) above. {\bf (c)} Same as in (b) above but for ${\cal{H}}_1$ prior
    pMSSM posterior samples. Here chargino-neutralino co-annihilations
    dominate.}
 \label{fig.hrLlogz}  
\end{figure}

The preferred content-nature of the neutralinos, based on the data and 
CDM prior assumption considered, can be seen from the pMSSM posterior 
samples. We use a measure of the gauginos content $Z_g = |N_{11}|^2 +
|N_{22}|^2$, so $1 -Z_g$ is approximately equal to unity or zero if
the neutralino is mostly higgsino- or gaugino-like respectively. The 
posterior distributions of $1-Z_g$ for both ${\cal{H}}_0$ (solid line)
and ${\cal{H}}_1$ (dashed line) pMSSM priors are shown in
Fig.~\ref{fig.hrLlogz}(a). It shows that the neutralino is mostly
higgsino-like for the ${\cal{H}}_1$ prior pMSSM. The distribution in the ${\cal{H}}_0$
prior is bimodal with dominantly gaugino-like neutralinos and
a sub-dominant mixed gaugino-higgsino peak. The drastic difference
between the posteriors resulting from the two hypotheses is due to
the fact that the CDM constraints imposed reject points that predict
relic densities much less than the observed central value in the
case of ${\cal{H}}_0$. As result the posterior samples are dominantly
gaugino-like but with sub-dominant mixed higgsino-gaugino neutralinos
that allow moderate co-annihilations that keep the relic density within
the ${\cal{H}}_0$ allowed range. ${\cal{H}}_1$ prior, on the other
hand, allow and in fact prefer (as it turned out to be) channels that
have very efficient co-annihilations which happen mostly for
higgsino-like neutralinos.  

The neutralino relic density is inversely proportional
to the thermally averaged magnitude of its annihilation and
co-annihilation reaction cross-sections at early universe time. The content-nature of the
neutralino determines which of the early universe reactions dominate
over others. For instance, a certain neutralino composition lead to an
acceptable magnitude for the thermally averaged cross-sections (which
is proportional to inverse of the relic density) with 8\% of the
cross-sections coming from direct $\tilde{\chi}^0_1$ annihilations
into W- 
and Z-bosons, 63\% from $\tilde{\chi}^0_1$-$\tilde{\chi}^0_2$
co-annihilations into quarks and 27\% into
leptons. Another example point has 46\% and 48\% of the 
thermally averaged cross-sections respectively coming from
$\tilde{\chi}^0_1$-$\tilde{u}$ and $\tilde{\chi}^0_1$-$\tilde{c}$
co-annihilations. We keep record of this information for each of the
pMSSM points visited during the nested sampling of the parameter
space. The posterior distributions for the dominant annihilation
and/or co-annihilation channels (or reactions) are shown in 
Fig.~\ref{fig.hrLlogz}(b) for ${\cal{H}}_0$ where the dominant channels
are co-annihilations with sleptons and in Fig.~\ref{fig.hrLlogz}(c)
for ${\cal{H}}_1$ where co-annihilations with charginos and annihilations via
chargino exchange dominate.   

\begin{figure}[t]
    \includegraphics[width=.9\textwidth, height=0.32\textwidth]{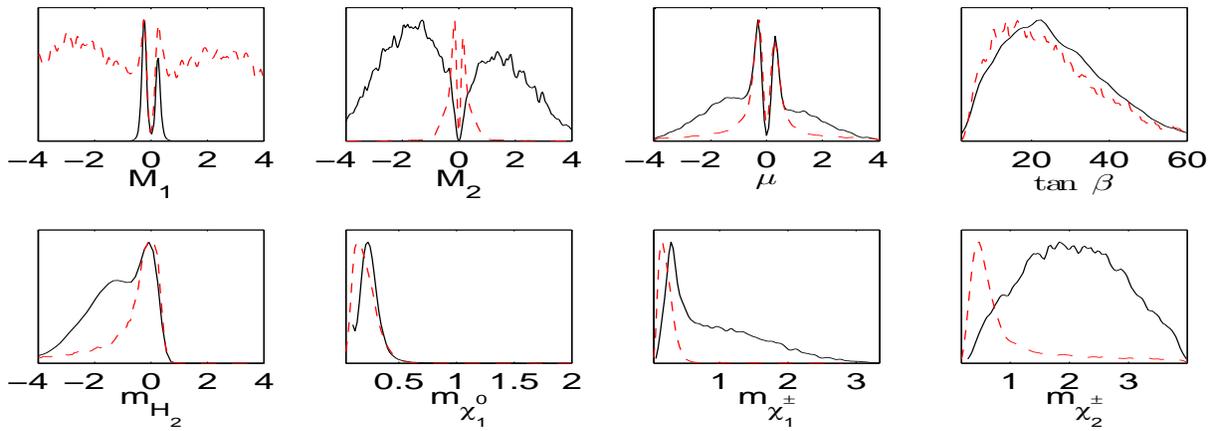}
  \caption{Marginalised one-dimensional posterior distributions for
    selected pMSSM parameters that are most affected by the choice
    between the 
    two hypotheses ${\cal{H}}_0$ and ${\cal{H}}_1$. Except for $\tan
    \beta$, all numbers in the horizontal axis are in TeV units. All
    the dashed lines are for the  
    prior hypothesis ${\cal{H}}_1$ where neutralinos form only part of
    multicomponent CDM. Solid lines are for the prior
    hypothesis ${\cal{H}}_0$ where neutralinos make up all of the CDM.} 
 \label{fig.hrLparams}
\end{figure}

The parameters posterior probability distributions are approximately the same for
both ${\cal{H}}_0$ and ${\cal{H}}_1$ pMSSM samples except in the neutralino and
chargino sectors. This is mainly due to the difference in the properties 
of the neutralinos that lead to different co-annihilation processes
which in turn control the consistency of the predicted relic densities
with the imposed CDM constraints. The neutralino mass, chargino mass
and the soft breaking parameters they depend on are shown in
Fig.~\ref{fig.hrLparams}. The neutralino mass and 
$\tan \beta$ are not severely constrained by the relic density
constraints and hence their posterior distributions do not
drastically differ between the hypotheses. The other
parameters and chargino masses, on the other hand, significantly
differ. The reasons for this are explained below. 
There is also a wide difference in the posterior probability distributions of the
SUSY breaking parameters $M_1$ and $M_2$ between the  
hypotheses since they determine the nature of the neutralino admixture
which is tightly related to the relic density predictions via the
preferred co-annihilation cross-sections. The $\mu$ parameter on the
other hand is not as widely different since it is mostly controlled by
the requirement of radiative electroweak symmetry breaking (EWSB) than
by the CDM constraint.   

The nature of the Higgs doublets mixing parameter, $\mu$, and mass
parameters $m_{H_{1,2}}$ are controlled at leading order by the tree
level requirement that  
\be \frac{1}{2} m_Z^2 = \frac{m_{{{H}}_1}^2 - m_{H_2}^2 \tan^2
  \beta}{\tan^2 \beta -1} -\mu^2
\ee 
at the EWSB scale. This reduces to 
$\frac{1}{2} m_Z^2 \simeq - m_{H_2}^2 - \mu^2$ with moderate to high
$\tan \beta$ values. That is for a fixed Z-boson mass, $m_{H_2}^2$ and
$\mu^2$ would be roughly the same since renormalisation group (RG)
running 
of the soft SUSY breaking parameters take $m_{H_2}$ to very large
(relative to the gaugino mass parameters) negative values at the EWSB
scale. This explains the shape of the 
diagrams for $\mu^2$ and $m_{H_2}^2$ parameters in 
Fig.~\ref{fig.hrLparams}. The ${\cal{H}}_1$ pMSSM $\mu^2$ posterior samples
(shown in dashed line in the figure) are less than those in the ${\cal{H}}_0$
prior because in the former case the neutralinos are mostly
higgsino-like compared to the dominantly gaugino-like neutralinos in
the latter. The details of the neutralino gaugino-higgsino content or
admixture is shown in Fig.~\ref{fig.hrLlogz}(a). 

The posterior distributions for the gaugino mass parameters $M_1$ and 
$M_2$ widely differs between the two CDM prior assumptions since these
parameters control the nature of the neutralino gaugino-higgsino
admixtures. We are going to address the $M_1$ and $M_2$ parameters
structure for each of the priors, starting with ${\cal{H}}_1$. The scenario in
the ${\cal{H}}_1$ prior is similar to the 
so-called focus point region in CMSSM where the $m_{H_2}$ RG
running is independent of the gaugino and trilinear parameters, see
e.g. \cite{Feng:1999zg}. 
As a result $M_1$ and $M_2$ would in principle remain unconstrained by
the EWSB requirement. This indeed remains the case for the $M_1$
parameter as shown in the corresponding plot in 
Fig.~\ref{fig.hrLparams}. However $M_2$ is constrained 
from the chargino sector since neutralino-chargino co-annihilations
requires $m_{\chi^\pm_1} \sim m_{\chi^0_1}$ which in turn imposes 
$M_2^2 \sim \mu^2$. The $M_2$ posterior
distribution is similar to that of the $\mu$ parameter as shown in
Fig.~\ref{fig.hrLparams} (dashed lines). Now turning to the ${\cal{H}}_0$
prior: again moderate $\tan \beta$ implies $m_{H_2}^2 \sim \mu^2$ but
here the neutralinos are mostly gauginos-like 
with $M_2^2 > \mu^2 > M_1^2$ and therefore different from the ${\cal{H}}_1$
focus point-like feature. $M_1$ is constrained to be small by
co-annihilation scenarios requiring mass of the neutralino, controlled
by $M_1 < M_2$, to be approximately degenerate with the sleptons it is
co-annihilating with. In summary $M_2$ controls the mass of the mostly
higgsino neutralinos via neutralino-chargino co-annihilations in ${\cal{H}}_1$
prior where the neutralinos make only part of the CDM energy of the
universe. On the other hand if neutralinos were to make all the CDM
then $M_1 < M_2$ controls its mass via co-annihilation with with
sleptons.   

We next turn to the chargino masses' posterior
distributions. Co-annihilations with charginos constrain the chargino
masses to be approximately equals those of the neutralinos. This is
more so in the case of the ${\cal{H}}_1$ prior (dashed lines in
Fig.~\ref{fig.hrLparams}) than in ${\cal{H}}_0$ prior (solid lines)
pMSSM since in the latter co-annihilations with sleptons dominate and
hence the chargino masses can be much larger than the neutralino
masses. The relation between $m_{\chi^\pm_1}$ and $m_{\chi^\pm_2}$ for 
each of the priors can be explained as follows. The masses
are controlled by the $M_2$ and $\mu$ parameters 
\be \label{char12}
m^2_{\chi^\pm_1, \chi^\pm_2} = \frac{1}{2} \left[ |M_2|^2 + |\mu|^2 +
  2m_W^2 \mp \sqrt{(|M_2|^2 + |\mu|^2 +  2m_W^2)^2 - 4 |\mu M_2 -m_W^2
    \sin 2\beta|^2} \right].
\ee
When $M_2^2 \sim \mu^2$, as is the case for the ${\cal{H}}_1$ prior pMSSM,
then the following approximation holds true $(|M_2|^2 + |\mu|^2 +
2m_W^2)^2 \sim 4 |\mu M_2 -m_W^2 \sin 2\beta|^2$. This in turn would
imply, from Eq.(\ref{char12}), that $m^2_{\chi^\pm_2} \sim
m^2_{\chi^\pm_1}$ as can be seen in the corresponding plots in
Fig.~\ref{fig.hrLparams}. In a similar manner when $M_2^2 >> \mu^2$ as
is the case for ${\cal{H}}_0$ prior pMSSM, then $\sqrt{(|M_2|^2 + |\mu|^2 +
  2m_W^2)^2 - 4 |\mu M_2 -m_W^2 \sin 2\beta|^2} \sim M_2^2.$ This way
$m^2_{\chi^\pm_1} \sim \frac{1}{2} \mu^2$ and $m^2_{\chi^\pm_2} \sim
M_2^2$ explain the structure of the corresponding ${\cal{H}}_0$ distributions
in Fig.~\ref{fig.hrLparams}.

Recapitulating, we have analysed 
the two mutually exclusive CDM hypotheses; whether it is solely made
of neutralinos, ${\cal{H}}_0$,  or whether a multicomponent scenario must
be allowed, ${\cal{H}}_1$. We applied the Bayesian approach to inference
and used the current indirect collider and cosmological data to select
between these hypotheses within the context of the pMSSM where $20$
phenomenologically viable MSSM parameters are simultaneously varied at
the electroweak energy scale. The pMSSM setting is unbiased and free
of theoretical assumptions 
and uncertainties from sources of SUSY breaking, mediation mechanisms
and RG running. We also
applied the Bayesian technique to  GUT-scale defined models of
SUSY mediation mechanisms. The results shown in
Tab.~\ref{tab:factors} give a data-based evidence that a
single neutralino CDM hypothesis is unrealistic and improbable. Based
on the set of data employed for the analysis, the probability that
neutralino make all the CDM for the pMSSM, mAMSB, CMSSM, and
LVS are about $0.06$ at most, $6.8 \times 10^{-7}$, $0.18$, and $0.02$
respectively. The results in Tab.~\ref{tab:factors} and Jeffrey's
scale (shown in Tab.~\ref{tab:Jeffreys}) for interpreting the evidence
values shows that prior dependence of the results are strongest in the
CMSSM, despite the fact that it has fewer number of parameters
compared to the pMSSM. CMSSM has a weak-to-moderate evidence
against ${\cal{H}}_0$, the solely neutralino made CDM hypothesis. The
pMSSM and LVS results have moderate-to-sub-strong evidence against
${\cal{H}}_0$. LVS 
result is approximately prior independent. The mAMSB model have an
exceptionally strong and prior independent evidence against ${\cal{H}}_0$. 
Based on these we therefore conclude that  MSSM
phenomenology studies should not neglect the mixed CDM possibility.  

Let us, here, re-emphasise that prior dependence is a positive feature
of Bayesian methods because it shows when the data and/or the physics
under consideration are strong or well-defined enough to be free of
ambiguous predictions and discriminations between alternative
hypotheses. Prior (in)dependence is not a 
feature simply determined by the number of parameters in the sense
that models with more parameters necessarily exhibit more prior
dependence in 
their predictions. We hope that the results presented in this paper
give good enough 
illustration of the mentioned observations. First, consider the mAMSB results in
Tab.~\ref{tab:factors}. The results are prior independent because the
physics is very well-pinned to the extent that even the
not-that-strong indirect collider and cosmological data,
Eq.(\ref{dat}), employed is good  
enough to give such exceptionally unambiguous results. Prior
dependence can be completely absent if the data employed is strong
enough. Secondly, the presence
of relatively stronger prior dependence in the model with less number
of parameters, CMSSM, compared to that with more parameters,
the pMSSM, illustrates that prior dependence is not
simply a feature directly derived from the number of model
parameters but is only a feature indicating the strength of the data
employed and/or the physics under consideration.

It is worth mentioning that it may  seem that the analysis is
not fair to the ${\cal{H}}_0$ prior pMSSM, where neutralinos are assumed to
form all CDM. Not fair in the sense that any additional CDM candidate would be accompanied by its own parameters describing 
its mass and couplings. It would seem that these additional
parameters would dilute the prior and in turn reduce the evidence for
the scenario. We emphasise that this is necessarily not going to be the case
because the additional DM parameters would make the model
under consideration something completely different from the pMSSM. As
such the comparison would then be between a pMSSM and a non-pMSSM. And
that is not what we considered in this article. The comparison carried
out is strictly between same pMSSMs with two distinct
and mutually exclusive hypotheses: in the first, ${\cal{H}}_0$, the CDM is assumed
to be solely made of the neutralino while in the second, ${\cal{H}}_1$, the
neutralino forms only part of the CDM with the remaining to
be accounted for by some other physics external and perpendicular to
the pMSSM. 

The posterior samples resulting from the analysis with ${\cal{H}}_0$, the
assumption that the neutralinos make all CDM, and ${\cal{H}}_1$, where the
neutralinos form only part of the CDM, hypotheses differ drastically
in the neutralino gaugino-higgsino content and neutralino/gauginos
sector parameters. The dark matter relic density constraint in ${\cal{H}}_0$ 
dis-favours pMSSM points with higgsino-like or mixed gaugino-higgsino
neutralinos that have efficient co-annihilation channels. However
relaxing the constraint to ${\cal{H}}_1$ 
disfavoured pMSSM points become the preferred ones. Results in both
this and previous pMSSM work indicate that, in light of the indirect
collider and cosmological data considered, neutralino-chargino
co-annihilations dominate at the early universe time before 
the neutralinos decouple from thermal equilibrium. It would be
interesting to see what implications this feature would have for MSSM
phenomenology at colliders.


Our results provide non-trivial quantitative evidence about the
composition of DM in the MSSM, even after having introduced
large number of parameters as in the pMSSM. We have assumed thermal
dark matter and R-parity conserving MSSM. Modification of these
assumptions or further extensions of the MSSM may modify our results.  
 
{\bf Acknowledgements:} We 
thank Ben Allanach, Alberto Casas and Matt Dolan for useful comments
and Mike Hobson for interesting discussion on Bayesian evidence
comparison for the neutralino CDM hypotheses. The large number of parameters involved in this project
required long sessions of high performance computing. We thank the University of Cambridge for 
direct access to the super- 
computers: COSMOS from the Department of Applied Mathematics and Theoretical 
Physics (DAMTP) and the Darwin cluster from the High Performance Computing 
Service (HPCS). 

\bibliography{references}

\begin{thebibliography}{70}
\expandafter\ifx\csname natexlab\endcsname\relax\def\natexlab#1{#1}\fi
\expandafter\ifx\csname bibnamefont\endcsname\relax
  \def\bibnamefont#1{#1}\fi
\expandafter\ifx\csname bibfnamefont\endcsname\relax
  \def\bibfnamefont#1{#1}\fi
\expandafter\ifx\csname citenamefont\endcsname\relax
  \def\citenamefont#1{#1}\fi
\expandafter\ifx\csname url\endcsname\relax
  \def\url#1{\texttt{#1}}\fi
\expandafter\ifx\csname urlprefix\endcsname\relax\def\urlprefix{URL }\fi
\providecommand{\bibinfo}[2]{#2}
\providecommand{\eprint}[2][]{\url{#2}}

\bibitem[{\citenamefont{Komatsu et~al.}(2010)}]{wmap7}
\bibinfo{author}{\bibfnamefont{E.}~\bibnamefont{Komatsu}} \bibnamefont{et~al.}
  (\bibinfo{year}{2010}), \eprint{1001.4538}.

\bibitem[{\citenamefont{Jungman et~al.}(1996)\citenamefont{Jungman,
  Kamionkowski, and Griest}}]{Jungman:1995df}
\bibinfo{author}{\bibfnamefont{G.}~\bibnamefont{Jungman}},
  \bibinfo{author}{\bibfnamefont{M.}~\bibnamefont{Kamionkowski}},
  \bibnamefont{and} \bibinfo{author}{\bibfnamefont{K.}~\bibnamefont{Griest}},
  \bibinfo{journal}{Phys. Rept.} \textbf{\bibinfo{volume}{267}},
  \bibinfo{pages}{195} (\bibinfo{year}{1996}), \eprint{hep-ph/9506380}.

\bibitem[{\citenamefont{Bertone et~al.}(2005)\citenamefont{Bertone, Hooper, and
  Silk}}]{Bertone:2004pz}
\bibinfo{author}{\bibfnamefont{G.}~\bibnamefont{Bertone}},
  \bibinfo{author}{\bibfnamefont{D.}~\bibnamefont{Hooper}}, \bibnamefont{and}
  \bibinfo{author}{\bibfnamefont{J.}~\bibnamefont{Silk}},
  \bibinfo{journal}{Phys. Rept.} \textbf{\bibinfo{volume}{405}},
  \bibinfo{pages}{279} (\bibinfo{year}{2005}), \eprint{hep-ph/0404175}.

\bibitem[{\citenamefont{Garrett and Duda}(2010)}]{Garrett:2010hd}
\bibinfo{author}{\bibfnamefont{K.}~\bibnamefont{Garrett}} \bibnamefont{and}
  \bibinfo{author}{\bibfnamefont{G.}~\bibnamefont{Duda}}
  (\bibinfo{year}{2010}), \eprint{1006.2483}.

\bibitem[{\citenamefont{Komatsu et~al.}(2009{\natexlab{a}})}]{arXiv:0803.0547}
\bibinfo{author}{\bibfnamefont{E.}~\bibnamefont{Komatsu}} \bibnamefont{et~al.}
  (\bibinfo{collaboration}{WMAP}), \bibinfo{journal}{Astrophys. J. Suppl.}
  \textbf{\bibinfo{volume}{180}}, \bibinfo{pages}{330}
  (\bibinfo{year}{2009}{\natexlab{a}}), \eprint{0803.0547}.

\bibitem[{\citenamefont{Giudice et~al.}(1998)\citenamefont{Giudice, Luty,
  Murayama, and Rattazzi}}]{Giudice:1998xp}
\bibinfo{author}{\bibfnamefont{G.~F.} \bibnamefont{Giudice}},
  \bibinfo{author}{\bibfnamefont{M.~A.} \bibnamefont{Luty}},
  \bibinfo{author}{\bibfnamefont{H.}~\bibnamefont{Murayama}}, \bibnamefont{and}
  \bibinfo{author}{\bibfnamefont{R.}~\bibnamefont{Rattazzi}},
  \bibinfo{journal}{JHEP} \textbf{\bibinfo{volume}{12}}, \bibinfo{pages}{027}
  (\bibinfo{year}{1998}), \eprint{hep-ph/9810442}.

\bibitem[{\citenamefont{Baer et~al.}(2010{\natexlab{a}})\citenamefont{Baer,
  Dermisek, Rajagopalan, and Summy}}]{Baer:2010kd}
\bibinfo{author}{\bibfnamefont{H.}~\bibnamefont{Baer}},
  \bibinfo{author}{\bibfnamefont{R.}~\bibnamefont{Dermisek}},
  \bibinfo{author}{\bibfnamefont{S.}~\bibnamefont{Rajagopalan}},
  \bibnamefont{and} \bibinfo{author}{\bibfnamefont{H.}~\bibnamefont{Summy}}
  (\bibinfo{year}{2010}{\natexlab{a}}), \eprint{1004.3297}.

\bibitem[{\citenamefont{Baer et~al.}(2010{\natexlab{b}})\citenamefont{Baer,
  Box, and Summy}}]{Baer:2010wm}
\bibinfo{author}{\bibfnamefont{H.}~\bibnamefont{Baer}},
  \bibinfo{author}{\bibfnamefont{A.~D.} \bibnamefont{Box}}, \bibnamefont{and}
  \bibinfo{author}{\bibfnamefont{H.}~\bibnamefont{Summy}}
  (\bibinfo{year}{2010}{\natexlab{b}}), \eprint{1005.2215}.

\bibitem[{\citenamefont{Conlon and Quevedo}(2007)}]{arXiv:0705.3460}
\bibinfo{author}{\bibfnamefont{J.~P.} \bibnamefont{Conlon}} \bibnamefont{and}
  \bibinfo{author}{\bibfnamefont{F.}~\bibnamefont{Quevedo}},
  \bibinfo{journal}{JCAP} \textbf{\bibinfo{volume}{0708}}, \bibinfo{pages}{019}
  (\bibinfo{year}{2007}), \eprint{0705.3460}.

\bibitem[{\citenamefont{Zurek}(2009)}]{Zurek:2008qg}
\bibinfo{author}{\bibfnamefont{K.~M.} \bibnamefont{Zurek}},
  \bibinfo{journal}{Phys. Rev.} \textbf{\bibinfo{volume}{D79}},
  \bibinfo{pages}{115002} (\bibinfo{year}{2009}), \eprint{0811.4429}.

\bibitem[{\citenamefont{Cholis and Weiner}(2009)}]{Cholis:2009va}
\bibinfo{author}{\bibfnamefont{I.}~\bibnamefont{Cholis}} \bibnamefont{and}
  \bibinfo{author}{\bibfnamefont{N.}~\bibnamefont{Weiner}}
  (\bibinfo{year}{2009}), \eprint{0911.4954}.

\bibitem[{\citenamefont{Feldman et~al.}(2010)\citenamefont{Feldman, Liu, Nath,
  and Peim}}]{Feldman:2010wy}
\bibinfo{author}{\bibfnamefont{D.}~\bibnamefont{Feldman}},
  \bibinfo{author}{\bibfnamefont{Z.}~\bibnamefont{Liu}},
  \bibinfo{author}{\bibfnamefont{P.}~\bibnamefont{Nath}}, \bibnamefont{and}
  \bibinfo{author}{\bibfnamefont{G.}~\bibnamefont{Peim}}
  (\bibinfo{year}{2010}), \eprint{1004.0649}.

\bibitem[{\citenamefont{Trotta}(2007)}]{astro-ph/0504022}
\bibinfo{author}{\bibfnamefont{R.}~\bibnamefont{Trotta}},
  \bibinfo{journal}{Mon. Not. Roy. Astron. Soc.}
  \textbf{\bibinfo{volume}{378}}, \bibinfo{pages}{72} (\bibinfo{year}{2007}),
  \eprint{astro-ph/0504022}.

\bibitem[{\citenamefont{von Weitershausen et~al.}(2010)\citenamefont{von
  Weitershausen, Schafer, Stockinger-Kim, and
  Stockinger}}]{vonWeitershausen:2010zr}
\bibinfo{author}{\bibfnamefont{P.}~\bibnamefont{von Weitershausen}},
  \bibinfo{author}{\bibfnamefont{M.}~\bibnamefont{Schafer}},
  \bibinfo{author}{\bibfnamefont{H.}~\bibnamefont{Stockinger-Kim}},
  \bibnamefont{and}
  \bibinfo{author}{\bibfnamefont{D.}~\bibnamefont{Stockinger}}
  (\bibinfo{year}{2010}), \eprint{1003.5820}.

\bibitem[{\citenamefont{Feroz et~al.}(2008{\natexlab{a}})}]{0807.4512}
\bibinfo{author}{\bibfnamefont{F.}~\bibnamefont{Feroz}} \bibnamefont{et~al.},
  \bibinfo{journal}{JHEP} \textbf{\bibinfo{volume}{10}}, \bibinfo{pages}{064}
  (\bibinfo{year}{2008}{\natexlab{a}}), \eprint{0807.4512}.

\bibitem[{\citenamefont{AbdusSalam et~al.}(2010)\citenamefont{AbdusSalam,
  Allanach, Quevedo, Feroz, and Hobson}}]{AbdusSalam:2009qd}
\bibinfo{author}{\bibfnamefont{S.~S.} \bibnamefont{AbdusSalam}},
  \bibinfo{author}{\bibfnamefont{B.~C.} \bibnamefont{Allanach}},
  \bibinfo{author}{\bibfnamefont{F.}~\bibnamefont{Quevedo}},
  \bibinfo{author}{\bibfnamefont{F.}~\bibnamefont{Feroz}}, \bibnamefont{and}
  \bibinfo{author}{\bibfnamefont{M.}~\bibnamefont{Hobson}},
  \bibinfo{journal}{Phys. Rev.} \textbf{\bibinfo{volume}{D81}},
  \bibinfo{pages}{095012} (\bibinfo{year}{2010}), \eprint{0904.2548}.

\bibitem[{\citenamefont{AbdusSalam et~al.}(2009)\citenamefont{AbdusSalam,
  Allanach, Dolan, Feroz, and Hobson}}]{AbdusSalam:2009tr}
\bibinfo{author}{\bibfnamefont{S.~S.} \bibnamefont{AbdusSalam}},
  \bibinfo{author}{\bibfnamefont{B.~C.} \bibnamefont{Allanach}},
  \bibinfo{author}{\bibfnamefont{M.~J.} \bibnamefont{Dolan}},
  \bibinfo{author}{\bibfnamefont{F.}~\bibnamefont{Feroz}}, \bibnamefont{and}
  \bibinfo{author}{\bibfnamefont{M.~P.} \bibnamefont{Hobson}},
  \bibinfo{journal}{Phys. Rev.} \textbf{\bibinfo{volume}{D80}},
  \bibinfo{pages}{035017} (\bibinfo{year}{2009}), \eprint{0906.0957}.

\bibitem[{\citenamefont{Allanach}(2006)}]{Allanach:2006jc}
\bibinfo{author}{\bibfnamefont{B.~C.} \bibnamefont{Allanach}},
  \bibinfo{journal}{Phys. Lett.} \textbf{\bibinfo{volume}{B635}},
  \bibinfo{pages}{123} (\bibinfo{year}{2006}), \eprint{hep-ph/0601089}.

\bibitem[{\citenamefont{Cabrera et~al.}(2009)\citenamefont{Cabrera, Casas, and
  Ruiz~de Austri}}]{Cabrera:2008tj}
\bibinfo{author}{\bibfnamefont{M.~E.} \bibnamefont{Cabrera}},
  \bibinfo{author}{\bibfnamefont{J.~A.} \bibnamefont{Casas}}, \bibnamefont{and}
  \bibinfo{author}{\bibfnamefont{R.}~\bibnamefont{Ruiz~de Austri}},
  \bibinfo{journal}{JHEP} \textbf{\bibinfo{volume}{03}}, \bibinfo{pages}{075}
  (\bibinfo{year}{2009}), \eprint{0812.0536}.

\bibitem[{\citenamefont{Raklev and White}(2009)}]{Raklev:2009uu}
\bibinfo{author}{\bibfnamefont{A.~R.} \bibnamefont{Raklev}} \bibnamefont{and}
  \bibinfo{author}{\bibfnamefont{M.~J.} \bibnamefont{White}}
  (\bibinfo{year}{2009}), \eprint{0911.1986}.

\bibitem[{\citenamefont{White and Feroz}(2010)}]{White:2010jp}
\bibinfo{author}{\bibfnamefont{M.~J.} \bibnamefont{White}} \bibnamefont{and}
  \bibinfo{author}{\bibfnamefont{F.}~\bibnamefont{Feroz}},
  \bibinfo{journal}{JHEP} \textbf{\bibinfo{volume}{07}}, \bibinfo{pages}{064}
  (\bibinfo{year}{2010}), \eprint{1002.1922}.

\bibitem[{\citenamefont{Djouadi et~al.}(1998)}]{Djouadi:1998di}
\bibinfo{author}{\bibfnamefont{A.}~\bibnamefont{Djouadi}} \bibnamefont{et~al.}
  (\bibinfo{collaboration}{MSSM Working Group}) (\bibinfo{year}{1998}),
  \eprint{hep-ph/9901246}.

\bibitem[{\citenamefont{AbdusSalam}(2009)}]{AbdusSalam:2008uv}
\bibinfo{author}{\bibfnamefont{S.~S.} \bibnamefont{AbdusSalam}},
  \bibinfo{journal}{AIP Conf. Proc.} \textbf{\bibinfo{volume}{1078}},
  \bibinfo{pages}{297} (\bibinfo{year}{2009}), \eprint{0809.0284}.

\bibitem[{\citenamefont{Nath et~al.}(2010)}]{Nath:2010zj}
\bibinfo{author}{\bibfnamefont{P.}~\bibnamefont{Nath}} \bibnamefont{et~al.},
  \bibinfo{journal}{Nucl. Phys. Proc. Suppl.}
  \textbf{\bibinfo{volume}{200-202}}, \bibinfo{pages}{185}
  (\bibinfo{year}{2010}), \eprint{1001.2693}.

\bibitem[{\citenamefont{{Skilling}}(2004)}]{Skilling}
\bibinfo{author}{\bibfnamefont{J.}~\bibnamefont{{Skilling}}}, in
  \emph{\bibinfo{booktitle}{American Institute of Physics Conference Series}},
  edited by \bibinfo{editor}{\bibfnamefont{R.}~\bibnamefont{{Fischer}}},
  \bibinfo{editor}{\bibfnamefont{R.}~\bibnamefont{{Preuss}}}, \bibnamefont{and}
  \bibinfo{editor}{\bibfnamefont{U.~V.} \bibnamefont{{Toussaint}}}
  (\bibinfo{year}{2004}), pp. \bibinfo{pages}{395--405},
  \urlprefix\url{http://www.inference.phy.cam.ac.uk/bayesys/}.

\bibitem[{\citenamefont{Feroz et~al.}(2008{\natexlab{b}})\citenamefont{Feroz,
  Hobson, and Bridges}}]{Feroz:2008xx}
\bibinfo{author}{\bibfnamefont{F.}~\bibnamefont{Feroz}},
  \bibinfo{author}{\bibfnamefont{M.~P.} \bibnamefont{Hobson}},
  \bibnamefont{and} \bibinfo{author}{\bibfnamefont{M.}~\bibnamefont{Bridges}},
  \bibinfo{journal}{Monthly Notices of the Royal Astronomical Society}
  \textbf{\bibinfo{volume}{398, 4}}, \bibinfo{pages}{1601}
  (\bibinfo{year}{2008}{\natexlab{b}}), \eprint{0809.3437}.

\bibitem[{\citenamefont{Feroz and Hobson}(2007)}]{Feroz:2007kg}
\bibinfo{author}{\bibfnamefont{F.}~\bibnamefont{Feroz}} \bibnamefont{and}
  \bibinfo{author}{\bibfnamefont{M.~P.} \bibnamefont{Hobson}},
  \bibinfo{journal}{Monthly Notices of the Royal Astronomical Society}
  \textbf{\bibinfo{volume}{384, 2}}, \bibinfo{pages}{449}
  (\bibinfo{year}{2007}), \eprint{0704.3704}.

\bibitem[{\citenamefont{{Jeffreys}}(1961)}]{Jeffreys}
\bibinfo{author}{\bibfnamefont{H.}~\bibnamefont{{Jeffreys}}},
  \emph{\bibinfo{title}{{Theory of probability, 3rd edn}}}
  (\bibinfo{publisher}{Oxford University Press}, \bibinfo{year}{1961}).

\bibitem[{\citenamefont{Berger et~al.}(2009)\citenamefont{Berger, Gainer,
  Hewett, and Rizzo}}]{Berger:2008cq}
\bibinfo{author}{\bibfnamefont{C.~F.} \bibnamefont{Berger}},
  \bibinfo{author}{\bibfnamefont{J.~S.} \bibnamefont{Gainer}},
  \bibinfo{author}{\bibfnamefont{J.~L.} \bibnamefont{Hewett}},
  \bibnamefont{and} \bibinfo{author}{\bibfnamefont{T.~G.} \bibnamefont{Rizzo}},
  \bibinfo{journal}{JHEP} \textbf{\bibinfo{volume}{02}}, \bibinfo{pages}{023}
  (\bibinfo{year}{2009}), \eprint{0812.0980}.

\bibitem[{\citenamefont{Baro et~al.}(2008)\citenamefont{Baro, Boudjema, and
  Semenov}}]{Baro:2007em}
\bibinfo{author}{\bibfnamefont{N.}~\bibnamefont{Baro}},
  \bibinfo{author}{\bibfnamefont{F.}~\bibnamefont{Boudjema}}, \bibnamefont{and}
  \bibinfo{author}{\bibfnamefont{A.}~\bibnamefont{Semenov}},
  \bibinfo{journal}{Phys. Lett.} \textbf{\bibinfo{volume}{B660}},
  \bibinfo{pages}{550} (\bibinfo{year}{2008}), \eprint{0710.1821}.

\bibitem[{\citenamefont{Allanach}(2002)}]{Allanach:2001kg}
\bibinfo{author}{\bibfnamefont{B.~C.} \bibnamefont{Allanach}},
  \bibinfo{journal}{Comput. Phys. Commun.} \textbf{\bibinfo{volume}{143}},
  \bibinfo{pages}{305} (\bibinfo{year}{2002}), \eprint{hep-ph/0104145}.

\bibitem[{\citenamefont{Belanger et~al.}(2008)\citenamefont{Belanger, Boudjema,
  Pukhov, and Semenov}}]{Belanger:2008sj}
\bibinfo{author}{\bibfnamefont{G.}~\bibnamefont{Belanger}},
  \bibinfo{author}{\bibfnamefont{F.}~\bibnamefont{Boudjema}},
  \bibinfo{author}{\bibfnamefont{A.}~\bibnamefont{Pukhov}}, \bibnamefont{and}
  \bibinfo{author}{\bibfnamefont{A.}~\bibnamefont{Semenov}}
  (\bibinfo{year}{2008}), \eprint{0803.2360}.

\bibitem[{\citenamefont{Mahmoudi}(2008)}]{Mahmoudi:2007vz}
\bibinfo{author}{\bibfnamefont{F.}~\bibnamefont{Mahmoudi}},
  \bibinfo{journal}{Comput. Phys. Commun.} \textbf{\bibinfo{volume}{178}},
  \bibinfo{pages}{745} (\bibinfo{year}{2008}), \eprint{0710.2067}.

\bibitem[{\citenamefont{Heinemeyer et~al.}(2006)\citenamefont{Heinemeyer,
  Hollik, Stockinger, Weber, and Weiglein}}]{Heinemeyer:2006px}
\bibinfo{author}{\bibfnamefont{S.}~\bibnamefont{Heinemeyer}},
  \bibinfo{author}{\bibfnamefont{W.}~\bibnamefont{Hollik}},
  \bibinfo{author}{\bibfnamefont{D.}~\bibnamefont{Stockinger}},
  \bibinfo{author}{\bibfnamefont{A.~M.} \bibnamefont{Weber}}, \bibnamefont{and}
  \bibinfo{author}{\bibfnamefont{G.}~\bibnamefont{Weiglein}},
  \bibinfo{journal}{JHEP} \textbf{\bibinfo{volume}{08}}, \bibinfo{pages}{052}
  (\bibinfo{year}{2006}), \eprint{hep-ph/0604147}.

\bibitem[{\citenamefont{Heinemeyer et~al.}(2008)\citenamefont{Heinemeyer,
  Hollik, Weber, and Weiglein}}]{Heinemeyer:2007bw}
\bibinfo{author}{\bibfnamefont{S.}~\bibnamefont{Heinemeyer}},
  \bibinfo{author}{\bibfnamefont{W.}~\bibnamefont{Hollik}},
  \bibinfo{author}{\bibfnamefont{A.~M.} \bibnamefont{Weber}}, \bibnamefont{and}
  \bibinfo{author}{\bibfnamefont{G.}~\bibnamefont{Weiglein}},
  \bibinfo{journal}{JHEP} \textbf{\bibinfo{volume}{04}}, \bibinfo{pages}{039}
  (\bibinfo{year}{2008}), \eprint{0710.2972}.

\bibitem[{\citenamefont{Verzocchi}(2008, Philadelphia, USA)}]{verzo}
\bibinfo{author}{\bibfnamefont{M.}~\bibnamefont{Verzocchi}}, in
  \emph{\bibinfo{booktitle}{{talk at ICHEP 2008}}} (\bibinfo{year}{2008,
  Philadelphia, USA}).

\bibitem[{\citenamefont{et~al.}()}]{spope}
\bibinfo{author}{\bibfnamefont{A.~M.~W.} \bibnamefont{et~al.}},
  \emph{\bibinfo{title}{SUSY-POPE (Precision Observables Precisely Evaluated)}}
  (\bibinfo{publisher}{in preparation}, ????).

\bibitem[{:20(2006)}]{:2005ema}
\bibinfo{journal}{Phys. Rept.} \textbf{\bibinfo{volume}{427}},
  \bibinfo{pages}{257} (\bibinfo{year}{2006}), \eprint{hep-ex/0509008}.

\bibitem[{\citenamefont{Moroi}(1996)}]{Moroi:1995yh}
\bibinfo{author}{\bibfnamefont{T.}~\bibnamefont{Moroi}},
  \bibinfo{journal}{Phys. Rev.} \textbf{\bibinfo{volume}{D53}},
  \bibinfo{pages}{6565} (\bibinfo{year}{1996}), \eprint{hep-ph/9512396}.

\bibitem[{\citenamefont{Degrassi and Giudice}(1998)}]{Degrassi:1998es}
\bibinfo{author}{\bibfnamefont{G.}~\bibnamefont{Degrassi}} \bibnamefont{and}
  \bibinfo{author}{\bibfnamefont{G.~F.} \bibnamefont{Giudice}},
  \bibinfo{journal}{Phys. Rev.} \textbf{\bibinfo{volume}{D58}},
  \bibinfo{pages}{053007} (\bibinfo{year}{1998}), \eprint{hep-ph/9803384}.

\bibitem[{\citenamefont{Heinemeyer
  et~al.}(2004{\natexlab{a}})\citenamefont{Heinemeyer, Stockinger, and
  Weiglein}}]{Heinemeyer:2003dq}
\bibinfo{author}{\bibfnamefont{S.}~\bibnamefont{Heinemeyer}},
  \bibinfo{author}{\bibfnamefont{D.}~\bibnamefont{Stockinger}},
  \bibnamefont{and} \bibinfo{author}{\bibfnamefont{G.}~\bibnamefont{Weiglein}},
  \bibinfo{journal}{Nucl. Phys.} \textbf{\bibinfo{volume}{B690}},
  \bibinfo{pages}{62} (\bibinfo{year}{2004}{\natexlab{a}}),
  \eprint{hep-ph/0312264}.

\bibitem[{\citenamefont{Heinemeyer
  et~al.}(2004{\natexlab{b}})\citenamefont{Heinemeyer, Stockinger, and
  Weiglein}}]{Heinemeyer:2004yq}
\bibinfo{author}{\bibfnamefont{S.}~\bibnamefont{Heinemeyer}},
  \bibinfo{author}{\bibfnamefont{D.}~\bibnamefont{Stockinger}},
  \bibnamefont{and} \bibinfo{author}{\bibfnamefont{G.}~\bibnamefont{Weiglein}},
  \bibinfo{journal}{Nucl. Phys.} \textbf{\bibinfo{volume}{B699}},
  \bibinfo{pages}{103} (\bibinfo{year}{2004}{\natexlab{b}}),
  \eprint{hep-ph/0405255}.

\bibitem[{\citenamefont{Bennett et~al.}(2006)}]{Bennett:2006fi}
\bibinfo{author}{\bibfnamefont{G.~W.} \bibnamefont{Bennett}}
  \bibnamefont{et~al.} (\bibinfo{collaboration}{Muon G-2}),
  \bibinfo{journal}{Phys. Rev.} \textbf{\bibinfo{volume}{D73}},
  \bibinfo{pages}{072003} (\bibinfo{year}{2006}), \eprint{hep-ex/0602035}.

\bibitem[{\citenamefont{Davier}(2007)}]{Davier:2007ua}
\bibinfo{author}{\bibfnamefont{M.}~\bibnamefont{Davier}},
  \bibinfo{journal}{Nucl. Phys. Proc. Suppl.} \textbf{\bibinfo{volume}{169}},
  \bibinfo{pages}{288} (\bibinfo{year}{2007}), \eprint{hep-ph/0701163}.

\bibitem[{\citenamefont{Hertzog et~al.}(2007)\citenamefont{Hertzog, Miller,
  de~Rafael, Lee~Roberts, and Stockinger}}]{Hertzog:2007hz}
\bibinfo{author}{\bibfnamefont{D.~W.} \bibnamefont{Hertzog}},
  \bibinfo{author}{\bibfnamefont{J.~P.} \bibnamefont{Miller}},
  \bibinfo{author}{\bibfnamefont{E.}~\bibnamefont{de~Rafael}},
  \bibinfo{author}{\bibfnamefont{B.}~\bibnamefont{Lee~Roberts}},
  \bibnamefont{and}
  \bibinfo{author}{\bibfnamefont{D.}~\bibnamefont{Stockinger}}
  (\bibinfo{year}{2007}), \eprint{0705.4617}.

\bibitem[{\citenamefont{Misiak et~al.}(2007)}]{Misiak:2006zs}
\bibinfo{author}{\bibfnamefont{M.}~\bibnamefont{Misiak}} \bibnamefont{et~al.},
  \bibinfo{journal}{Phys. Rev. Lett.} \textbf{\bibinfo{volume}{98}},
  \bibinfo{pages}{022002} (\bibinfo{year}{2007}), \eprint{hep-ph/0609232}.

\bibitem[{\citenamefont{Misiak}(2006)}]{Misiak:2006bw}
\bibinfo{author}{\bibfnamefont{M.}~\bibnamefont{Misiak}}
  (\bibinfo{year}{2006}), \eprint{hep-ph/0609289}.

\bibitem[{\citenamefont{Misiak and Steinhauser}(2007)}]{Misiak:2006ab}
\bibinfo{author}{\bibfnamefont{M.}~\bibnamefont{Misiak}} \bibnamefont{and}
  \bibinfo{author}{\bibfnamefont{M.}~\bibnamefont{Steinhauser}},
  \bibinfo{journal}{Nucl. Phys.} \textbf{\bibinfo{volume}{B764}},
  \bibinfo{pages}{62} (\bibinfo{year}{2007}), \eprint{hep-ph/0609241}.

\bibitem[{\citenamefont{Becher and Neubert}(2007)}]{Becher:2006pu}
\bibinfo{author}{\bibfnamefont{T.}~\bibnamefont{Becher}} \bibnamefont{and}
  \bibinfo{author}{\bibfnamefont{M.}~\bibnamefont{Neubert}},
  \bibinfo{journal}{Phys. Rev. Lett.} \textbf{\bibinfo{volume}{98}},
  \bibinfo{pages}{022003} (\bibinfo{year}{2007}), \eprint{hep-ph/0610067}.

\bibitem[{\citenamefont{Barberio et~al.}(2007)}]{Barberio:2007cr}
\bibinfo{author}{\bibfnamefont{E.}~\bibnamefont{Barberio}} \bibnamefont{et~al.}
  (\bibinfo{collaboration}{Heavy Flavor Averaging Group (HFAG)})
  (\bibinfo{year}{2007}), \eprint{0704.3575}.

\bibitem[{\citenamefont{Belanger et~al.}(2007)\citenamefont{Belanger, Boudjema,
  Pukhov, and Semenov}}]{Belanger:2006is}
\bibinfo{author}{\bibfnamefont{G.}~\bibnamefont{Belanger}},
  \bibinfo{author}{\bibfnamefont{F.}~\bibnamefont{Boudjema}},
  \bibinfo{author}{\bibfnamefont{A.}~\bibnamefont{Pukhov}}, \bibnamefont{and}
  \bibinfo{author}{\bibfnamefont{A.}~\bibnamefont{Semenov}},
  \bibinfo{journal}{Comput. Phys. Commun.} \textbf{\bibinfo{volume}{176}},
  \bibinfo{pages}{367} (\bibinfo{year}{2007}), \eprint{hep-ph/0607059}.

\bibitem[{\citenamefont{Belanger et~al.}(2006)\citenamefont{Belanger, Boudjema,
  Pukhov, and Semenov}}]{Belanger:2004yn}
\bibinfo{author}{\bibfnamefont{G.}~\bibnamefont{Belanger}},
  \bibinfo{author}{\bibfnamefont{F.}~\bibnamefont{Boudjema}},
  \bibinfo{author}{\bibfnamefont{A.}~\bibnamefont{Pukhov}}, \bibnamefont{and}
  \bibinfo{author}{\bibfnamefont{A.}~\bibnamefont{Semenov}},
  \bibinfo{journal}{Comput. Phys. Commun.} \textbf{\bibinfo{volume}{174}},
  \bibinfo{pages}{577} (\bibinfo{year}{2006}), \eprint{hep-ph/0405253}.

\bibitem[{\citenamefont{Belanger et~al.}(2002)\citenamefont{Belanger, Boudjema,
  Pukhov, and Semenov}}]{Belanger:2001fz}
\bibinfo{author}{\bibfnamefont{G.}~\bibnamefont{Belanger}},
  \bibinfo{author}{\bibfnamefont{F.}~\bibnamefont{Boudjema}},
  \bibinfo{author}{\bibfnamefont{A.}~\bibnamefont{Pukhov}}, \bibnamefont{and}
  \bibinfo{author}{\bibfnamefont{A.}~\bibnamefont{Semenov}},
  \bibinfo{journal}{Comput. Phys. Commun.} \textbf{\bibinfo{volume}{149}},
  \bibinfo{pages}{103} (\bibinfo{year}{2002}), \eprint{hep-ph/0112278}.

\bibitem[{\citenamefont{Aaltonen et~al.}(2008)}]{:2007kv}
\bibinfo{author}{\bibfnamefont{T.}~\bibnamefont{Aaltonen}} \bibnamefont{et~al.}
  (\bibinfo{collaboration}{CDF}), \bibinfo{journal}{Phys. Rev. Lett.}
  \textbf{\bibinfo{volume}{100}}, \bibinfo{pages}{101802}
  (\bibinfo{year}{2008}), \eprint{0712.1708}.

\bibitem[{\citenamefont{Bona et~al.}(2006)}]{Bona:2006ah}
\bibinfo{author}{\bibfnamefont{M.}~\bibnamefont{Bona}} \bibnamefont{et~al.}
  (\bibinfo{collaboration}{UTfit}), \bibinfo{journal}{JHEP}
  \textbf{\bibinfo{volume}{10}}, \bibinfo{pages}{081} (\bibinfo{year}{2006}),
  \eprint{hep-ph/0606167}.

\bibitem[{\citenamefont{Abulencia et~al.}(2006)}]{Abulencia:2006ze}
\bibinfo{author}{\bibfnamefont{A.}~\bibnamefont{Abulencia}}
  \bibnamefont{et~al.} (\bibinfo{collaboration}{CDF}), \bibinfo{journal}{Phys.
  Rev. Lett.} \textbf{\bibinfo{volume}{97}}, \bibinfo{pages}{242003}
  (\bibinfo{year}{2006}), \eprint{hep-ex/0609040}.

\bibitem[{\citenamefont{Isidori and Paradisi}(2006)}]{Isidori:2006pk}
\bibinfo{author}{\bibfnamefont{G.}~\bibnamefont{Isidori}} \bibnamefont{and}
  \bibinfo{author}{\bibfnamefont{P.}~\bibnamefont{Paradisi}},
  \bibinfo{journal}{Phys. Lett.} \textbf{\bibinfo{volume}{B639}},
  \bibinfo{pages}{499} (\bibinfo{year}{2006}), \eprint{hep-ph/0605012}.

\bibitem[{\citenamefont{Isidori et~al.}(2007)\citenamefont{Isidori, Mescia,
  Paradisi, and Temes}}]{Isidori:2007jw}
\bibinfo{author}{\bibfnamefont{G.}~\bibnamefont{Isidori}},
  \bibinfo{author}{\bibfnamefont{F.}~\bibnamefont{Mescia}},
  \bibinfo{author}{\bibfnamefont{P.}~\bibnamefont{Paradisi}}, \bibnamefont{and}
  \bibinfo{author}{\bibfnamefont{D.}~\bibnamefont{Temes}},
  \bibinfo{journal}{Phys. Rev.} \textbf{\bibinfo{volume}{D75}},
  \bibinfo{pages}{115019} (\bibinfo{year}{2007}), \eprint{hep-ph/0703035}.

\bibitem[{\citenamefont{Akeroyd and Recksiegel}(2003)}]{Akeroyd:2003zr}
\bibinfo{author}{\bibfnamefont{A.~G.} \bibnamefont{Akeroyd}} \bibnamefont{and}
  \bibinfo{author}{\bibfnamefont{S.}~\bibnamefont{Recksiegel}},
  \bibinfo{journal}{J. Phys.} \textbf{\bibinfo{volume}{G29}},
  \bibinfo{pages}{2311} (\bibinfo{year}{2003}), \eprint{hep-ph/0306037}.

\bibitem[{\citenamefont{Aubert et~al.}(2005)}]{Aubert:2004kz}
\bibinfo{author}{\bibfnamefont{B.}~\bibnamefont{Aubert}} \bibnamefont{et~al.}
  (\bibinfo{collaboration}{BABAR}), \bibinfo{journal}{Phys. Rev. Lett.}
  \textbf{\bibinfo{volume}{95}}, \bibinfo{pages}{041804}
  (\bibinfo{year}{2005}), \eprint{hep-ex/0407038}.

\bibitem[{\citenamefont{Chang}(2008, Philadelphia, USA)}]{paoti}
\bibinfo{author}{\bibfnamefont{P.}~\bibnamefont{Chang}}, in
  \emph{\bibinfo{booktitle}{{talk at ICHEP 2008}}} (\bibinfo{year}{2008,
  Philadelphia, USA}).

\bibitem[{\citenamefont{Gray et~al.}(2005)}]{hep-lat/0507015}
\bibinfo{author}{\bibfnamefont{A.}~\bibnamefont{Gray}} \bibnamefont{et~al.}
  (\bibinfo{collaboration}{HPQCD}), \bibinfo{journal}{Phys. Rev. Lett.}
  \textbf{\bibinfo{volume}{95}}, \bibinfo{pages}{212001}
  (\bibinfo{year}{2005}), \eprint{hep-lat/0507015}.

\bibitem[{\citenamefont{Amsler et~al.}(2008)}]{J.Phys.G33.1}
\bibinfo{author}{\bibfnamefont{C.}~\bibnamefont{Amsler}} \bibnamefont{et~al.}
  (\bibinfo{collaboration}{Particle Data Group}), \bibinfo{journal}{Phys.
  Lett.} \textbf{\bibinfo{volume}{B667}}, \bibinfo{pages}{1}
  (\bibinfo{year}{2008}).

\bibitem[{\citenamefont{Komatsu et~al.}(2009{\natexlab{b}})}]{0803.0547}
\bibinfo{author}{\bibfnamefont{E.}~\bibnamefont{Komatsu}} \bibnamefont{et~al.}
  (\bibinfo{collaboration}{WMAP}), \bibinfo{journal}{Astrophys. J. Suppl.}
  \textbf{\bibinfo{volume}{180}}, \bibinfo{pages}{330}
  (\bibinfo{year}{2009}{\natexlab{b}}), \eprint{0803.0547}.

\bibitem[{\citenamefont{Conlon et~al.}(2007)\citenamefont{Conlon, Abdussalam,
  Quevedo, and Suruliz}}]{Conlon:2006wz}
\bibinfo{author}{\bibfnamefont{J.~P.} \bibnamefont{Conlon}},
  \bibinfo{author}{\bibfnamefont{S.~S.} \bibnamefont{Abdussalam}},
  \bibinfo{author}{\bibfnamefont{F.}~\bibnamefont{Quevedo}}, \bibnamefont{and}
  \bibinfo{author}{\bibfnamefont{K.}~\bibnamefont{Suruliz}},
  \bibinfo{journal}{JHEP} \textbf{\bibinfo{volume}{01}}, \bibinfo{pages}{032}
  (\bibinfo{year}{2007}), \eprint{hep-th/0610129}.

\bibitem[{\citenamefont{AbdusSalam et~al.}(2007)\citenamefont{AbdusSalam,
  Conlon, Quevedo, and Suruliz}}]{AbdusSalam:2007pm}
\bibinfo{author}{\bibfnamefont{S.~S.} \bibnamefont{AbdusSalam}},
  \bibinfo{author}{\bibfnamefont{J.~P.} \bibnamefont{Conlon}},
  \bibinfo{author}{\bibfnamefont{F.}~\bibnamefont{Quevedo}}, \bibnamefont{and}
  \bibinfo{author}{\bibfnamefont{K.}~\bibnamefont{Suruliz}},
  \bibinfo{journal}{JHEP} \textbf{\bibinfo{volume}{12}}, \bibinfo{pages}{036}
  (\bibinfo{year}{2007}), \eprint{0709.0221}.

\bibitem[{\citenamefont{El~Kheishen et~al.}(1992)\citenamefont{El~Kheishen,
  Aboshousha, and Shafik}}]{ElKheishen:1992yv}
\bibinfo{author}{\bibfnamefont{M.~M.} \bibnamefont{El~Kheishen}},
  \bibinfo{author}{\bibfnamefont{A.~A.} \bibnamefont{Aboshousha}},
  \bibnamefont{and} \bibinfo{author}{\bibfnamefont{A.~A.}
  \bibnamefont{Shafik}}, \bibinfo{journal}{Phys. Rev.}
  \textbf{\bibinfo{volume}{D45}}, \bibinfo{pages}{4345} (\bibinfo{year}{1992}).

\bibitem[{\citenamefont{Feng et~al.}(2000)\citenamefont{Feng, Matchev, and
  Moroi}}]{Feng:1999zg}
\bibinfo{author}{\bibfnamefont{J.~L.} \bibnamefont{Feng}},
  \bibinfo{author}{\bibfnamefont{K.~T.} \bibnamefont{Matchev}},
  \bibnamefont{and} \bibinfo{author}{\bibfnamefont{T.}~\bibnamefont{Moroi}},
  \bibinfo{journal}{Phys. Rev.} \textbf{\bibinfo{volume}{D61}},
  \bibinfo{pages}{075005} (\bibinfo{year}{2000}), \eprint{hep-ph/9909334}.

\bibitem[{\citenamefont{Lewin and Smith}(1996)}]{Lewin199687}
\bibinfo{author}{\bibfnamefont{J.~D.} \bibnamefont{Lewin}} \bibnamefont{and}
  \bibinfo{author}{\bibfnamefont{P.~F.} \bibnamefont{Smith}},
  \bibinfo{journal}{Astroparticle Physics} \textbf{\bibinfo{volume}{6}},
  \bibinfo{pages}{87 } (\bibinfo{year}{1996}), ISSN \bibinfo{issn}{0927-6505}.

\bibitem[{\citenamefont{Buchmueller et~al.}(2008)}]{Buchmueller:2008qe}
\bibinfo{author}{\bibfnamefont{O.}~\bibnamefont{Buchmueller}}
  \bibnamefont{et~al.}, \bibinfo{journal}{JHEP} \textbf{\bibinfo{volume}{09}},
  \bibinfo{pages}{117} (\bibinfo{year}{2008}), \eprint{0808.4128}.

\end{thebibliography}
\end{document}